\pdfoutput=1

\documentclass[aps,prl,twocolumn,superscriptaddress,longbibliography]{revtex4-1}

\usepackage[normalem]{ulem}
\usepackage{caption}
\usepackage{mathrsfs}
\usepackage{nicefrac}
\usepackage{amssymb}
\usepackage{eqnarray}
\usepackage{graphicx,bm,amsmath}
\usepackage[usenames,dvipsnames]{color}
\usepackage[colorlinks,bookmarks=false,citecolor=NavyBlue,linkcolor=Red,urlcolor=blue]{hyperref}
\usepackage[bbgreekl]{mathbbol}
\usepackage{xcolor}
\usepackage{pgf}
\usepackage{comment}
\usepackage{subfig}
\usepackage{rotating}

\usepackage{bbold}
\newcommand{\Id}{{\mathbb 1}}
\newcommand{\Ha}{\mathcal{H}}	
\newcommand{\DLayer}{\mathcal{D}(u)}
\newcommand{\DLayerd}{\mathcal{D}^\dagger(u)}
\newcommand{\Hilb}{\mathscr{H}}
\newcommand{\Lat}{\mathscr{L}}

\renewcommand{\vec}[1]{\mathbf{#1}}
\DeclareMathOperator{\tr}{tr}

\newcommand{\bra}[1]{\left\langle#1\right|}
\newcommand{\ket}[1]{\left|#1\right\rangle}

\newcommand{\be}{\begin{equation}}
\newcommand{\ee}{\end{equation}}
\newcommand{\bea}{\begin{eqnarray}}
\newcommand{\eea}{\end{eqnarray}}

\newcommand{\nep}{\textrm{e}}
\definecolor{smoothred}{HTML}{C5232F}
\definecolor{mygreen}{rgb}{0,0.5,0}
\definecolor{myblue}{rgb}{0,0,0.75}
\definecolor{mymagenta}{cmyk}{0,1,0,0.12}

\def\doi{http://dx.doi.org/}

\begin{document}

\title{Efficient Tensor Network ansatz for high-dimensional quantum many-body problems}

\author{Timo Felser}
\affiliation{Theoretische Physik, Universit\"at des Saarlandes, D-66123 Saarbr\"ucken, Germany.}
\affiliation{Dipartimento di Fisica e Astronomia ``G. Galilei'', Universit\`a di Padova, I-35131 Padova, Italy}
\affiliation{INFN, Sezione di Padova, Via Marzolo 8, I-35131 Padova, Italy}   

\author{Simone Notarnicola}
\affiliation{Dipartimento di Fisica e Astronomia ``G. Galilei'', Universit\`a di Padova, I-35131 Padova, Italy}
\affiliation{INFN, Sezione di Padova, Via Marzolo 8, I-35131 Padova, Italy}   

\author{Simone Montangero}
\affiliation{Dipartimento di Fisica e Astronomia ``G. Galilei'', Universit\`a di Padova, I-35131 Padova, Italy}   
\affiliation{INFN, Sezione di Padova, Via Marzolo 8, I-35131 Padova, Italy}

\date{\today}

\begin{abstract}
We introduce a novel tensor network structure augmenting the well-established Tree Tensor Network representation of a quantum many-body wave function. The new structure satisfies the area law in high dimensions remaining efficiently manipulatable and scalable. We benchmark this novel approach against paradigmatic two-dimensional spin models demonstrating unprecedented precision and system sizes. Finally, we compute the ground state phase diagram of  two-dimensional lattice Rydberg atoms in optical tweezers observing non-trivial phases and quantum phase transitions, providing realistic benchmarks for current and future two-dimensional quantum simulations. 
\end{abstract}

\maketitle

\begin{figure}[t!]
    \pgfputat{\pgfxy(0.0,-2.0)}{\pgfbox[left,bottom]{\footnotesize  \textcolor{RoyalBlue}{$ |\psi_{TTN}\rangle$}}}
   \pgfputat{\pgfxy(0.41,-2.7)}{\pgfbox[left,bottom]{\footnotesize  \textcolor{OliveGreen}{$ \DLayer$}}}
   \pgfputat{\pgfxy(0.61,-3.1)}{\pgfbox[left,bottom]{\footnotesize \textcolor{Bittersweet}{$ \Ha$}}}
   \pgfputat{\pgfxy(0.0,-0.3)}{\pgfbox[left,bottom]{\footnotesize (a)}}
   \pgfputat{\pgfxy(4.4,-0.3)}{\pgfbox[left,bottom]{\footnotesize (b)}}
   \pgfputat{\pgfxy(4.4,-4.2)}{\pgfbox[left,bottom]{\footnotesize (d)}}
   \pgfputat{\pgfxy(0.0,-4.2)}{\pgfbox[left,bottom]{\footnotesize (c)}}
   \pgfputat{\pgfxy(5.5,-0.6)}{\pgfbox[left,bottom]{\footnotesize Area law in 2D}}
\begin{tabular}{cc}
& \\
~~~~~~ \includegraphics[width=0.16\textwidth]{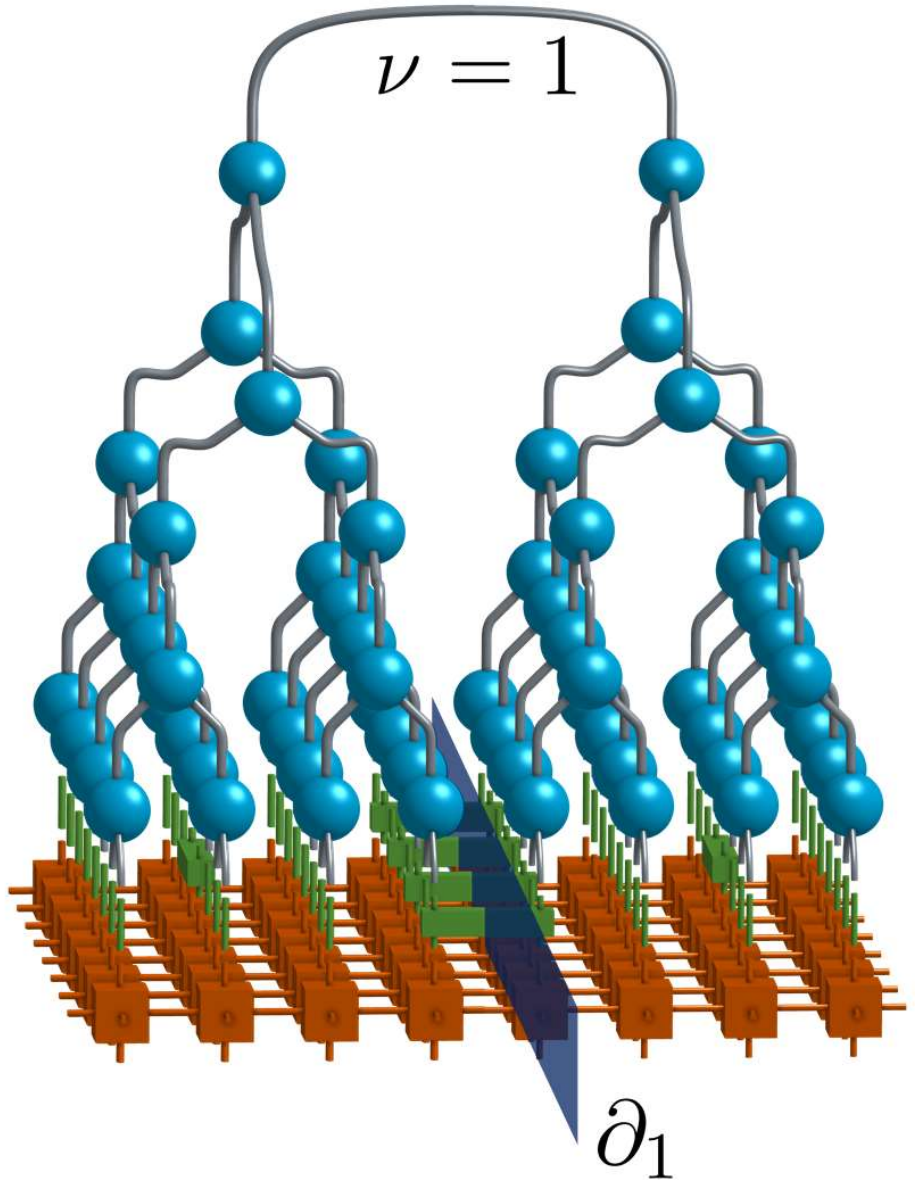} &
~~~~\includegraphics[width=0.19\textwidth]{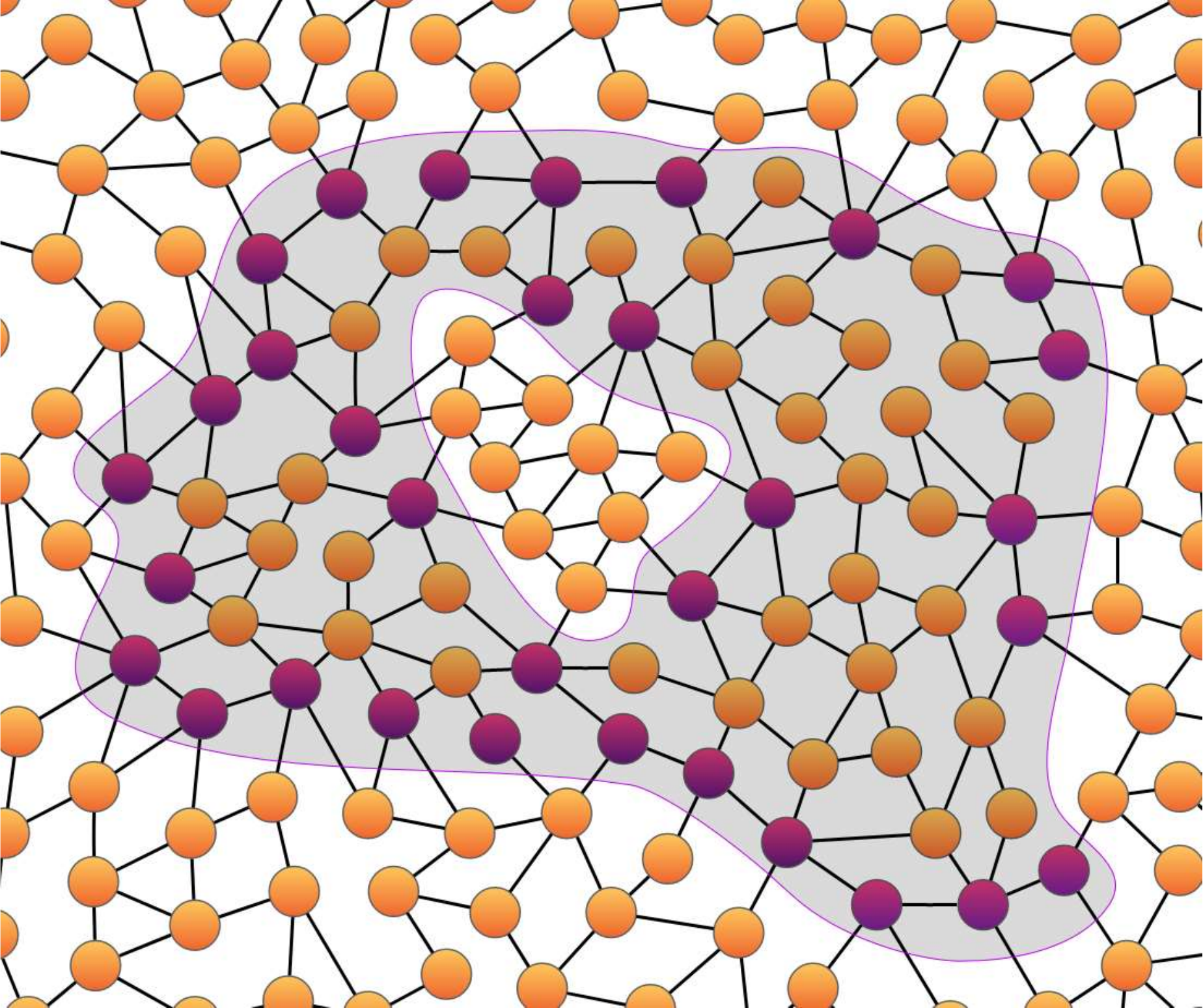} \\ & \\
\includegraphics[width=0.23\textwidth]{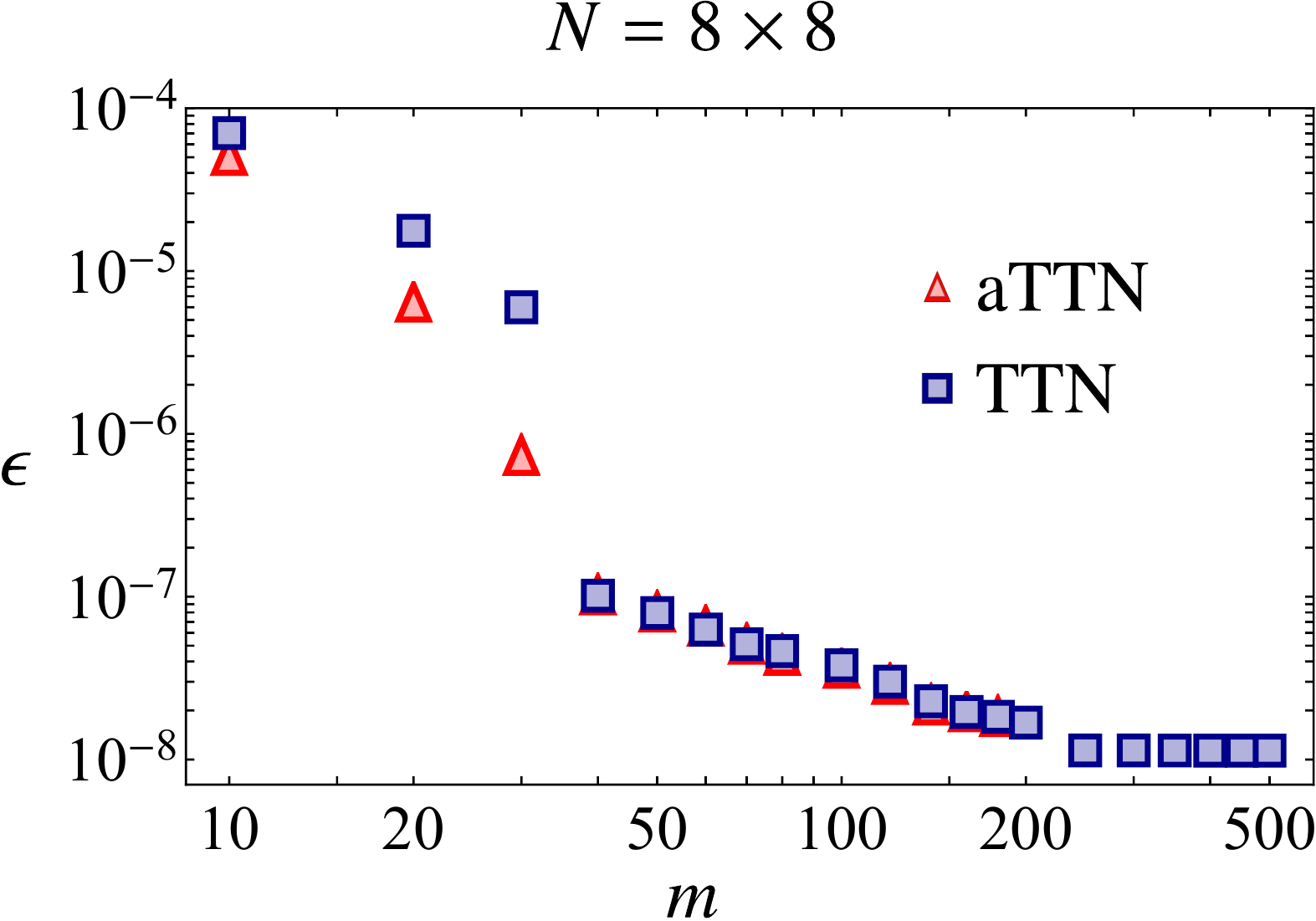} &
\includegraphics[width=0.23\textwidth]{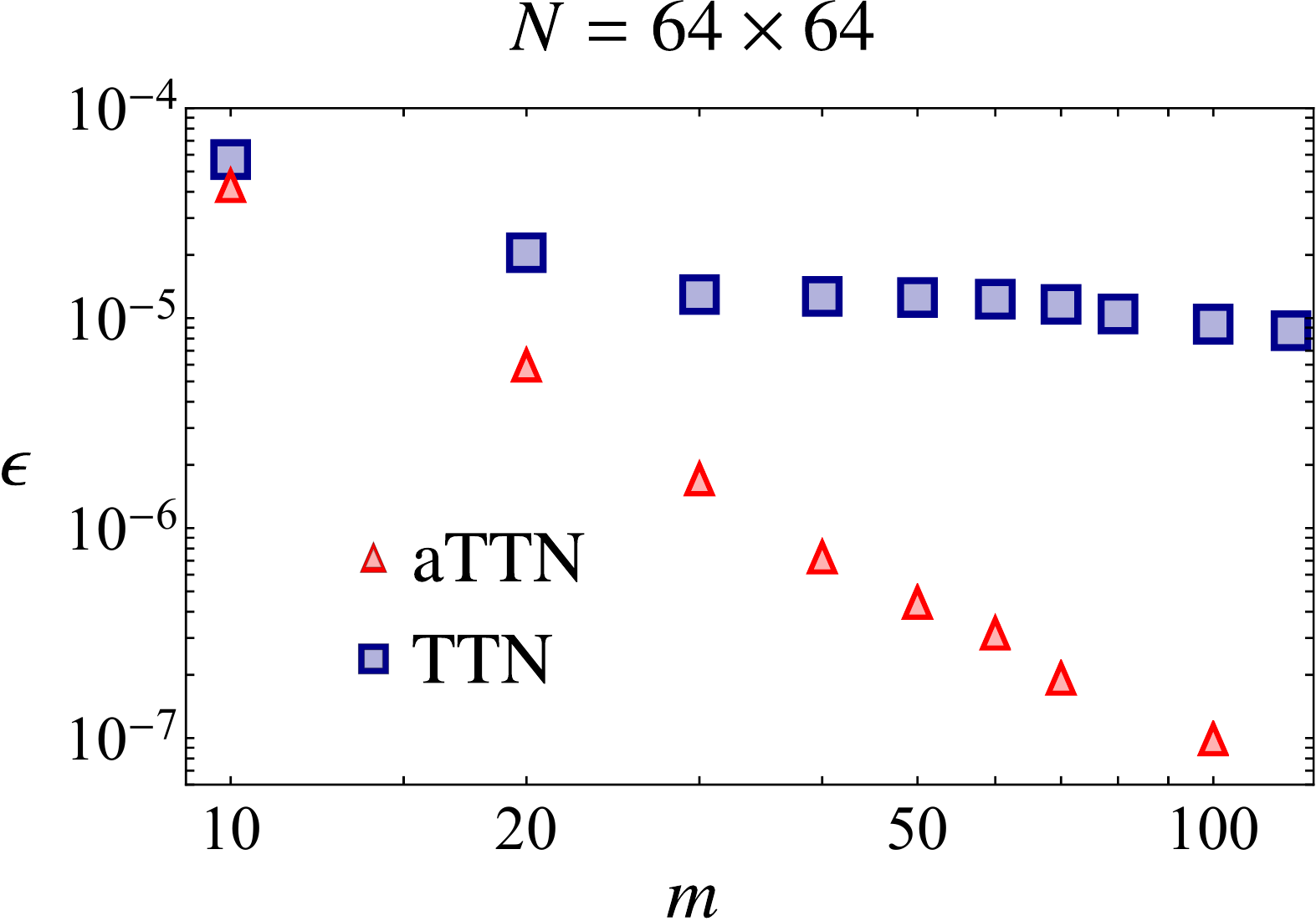}
\end{tabular}
\captionsetup{justification=centerlast}
\caption{\label{fig:intro}
(a) An aTTN for a $8\times 8$ 2D system: The disentanglers in $\DLayer$ are applied to the TTN state $ |\psi_{TTN}\rangle$ across the boundaries $\partial_\nu$ of each link $\nu$, in order to fulfill the area law depicted (b) for a sublattice $\mathcal{A}$ (shaded region) and its boundary $\partial \mathcal{A}$ (purple dots).(c), (d): relative error of the Ising model ground state energy computed with the aTTNs and the TTNs. While for $L=8$ the precision achieved with the two methods is the same, a clear improvement emerges for $L=64$.
}
\end{figure}

Recent experiments investigated one- and two-dimensional lattice quantum many-body systems at unprecedented sizes, calling for a continuous search of numerical techniques to provide accurate benchmarking and verification of future quantum simulations \cite{deLesleuc775,Browaeys2020,Rui2020,Bernien2017,Schauss2012,PhysRevLett.124.010504,2020arXiv200811389O,Kokail2019,PhysRevX.10.021057}. In particular, Rydberg atoms in optical tweezers are one of the most promising platforms for the study of quantum phase transitions, quantum simulation and computation~\cite{Labuhn2016,Schauss_2018,PhysRevResearch.2.013288,PhysRevX.10.021041,Barredo2018,PhysRevLett.124.023201,PhysRevX.8.021070,Omran570,PhysRevLett.123.170503,2020arXiv201103031M}. 
In the last decades, Monte Carlo and Tensor Networks (TN) algorithms have been employed widely to study quantum many-body systems, and they are routinely used to benchmark quantum simulation results~\cite{Koloren__2011,ACIOLI199775,Sandvik_1997,SimoneBook,felser2019twodimensional,Orus2019,Arute2019,Banuls2020,PhysRevA.101.023617,PhysRevResearch.2.013145,PhysRevD.102.074501}. However, Monte Carlo methods are limited by the sign problem~\cite{PhysRevLett.94.170201}, while combining accuracy and scalability in simulating high-dimensional systems still represent an open challenge for TN methods~\cite{TTNA19,Ran2020}. 
Here, we introduce a novel TN variational ansatz, able to encode the area law of quantum many-body states in any spatial dimension { by keeping a low} algorithmic complexity with respect to standard algorithms (see Fig.~\ref{fig:intro}), thus opening a pathway towards the application of TN to high-dimensional systems. Hereafter, we benchmark this approach against spin models up to sizes of $N=64\times 64$, in and out of criticality. Finally, we simulate 2D lattices of $N\sim 1000$ Rydberg atoms obtaining a phase diagram which exhibits nontrivial phase transitions, complementing recent results concerning a quasi two-dimensional similar model \cite{PhysRevLett.124.103601}.

In the last three decades, TN have been developed and applied to classically simulate quantum many-body systems,  representing the exponentially large wavefunction with a set of local tensors connected via auxiliary indices with a \textit{bond-dimension} $m$. The bond dimension $m$ allows to control the amount of  information in the TN,  interpolating between mean field ($m=1$) and the exact but inefficient representation. 
While for one-dimensional systems the Matrix Product States (MPS) are the established TN geometry for equilibrium and out-of-equilibrium problems, the development of TN algorithms for two- or even higher-dimensional systems is still ongoing~\cite{OstlundRommer,MPSfaithful,WhiteDMRG92,DMRGageofMPS,TDMRG,TEBD}. The most successful TN representations are the Projected Entangled Pair States (PEPS)~\cite{PEPSfirsttime2004, PEPSbasics2006,OrusPEPS, verstraete2004renormalization} 
and the Tree Tensor Networks (TTN)~\cite{VidalDuanTTN2006,HomogeneousTTN,TTN14,gqh17}, as well as the Multi-scale Entanglement
Renormalization Ansatz (MERA)~\cite{MERAfirst,MERA2D,MERA_bible}.
 %%%%%% SKETCH OF THE ATTN GEOMETRY %%%%%%%%%%%%
%%%%%%%%%%%%%%%%%%%%%%%%%%%%%%%%%%%%%%%%%%%%%%%:

{TNs shall satisfy the same entanglement bounds under real-space bipartitions, known as area laws, of the \textit{physical states} they represent~\cite{PEPSbasics2006, MPSfaithful, Eisert_2010}.
The PEPS is the potentially most powerful TN ansatz and by construction satisfies the area laws of entanglement~\cite{Eisert_2010}. 
However, it suffers from a high algorithmical complexity ($\mathcal{O}\left(m^{10}\right)$) and lacks an exact calculation of expectation values. Indeed, the exact contraction a finite square lattice of the complete PEPS scales exponentially with the system linear dimension %on average system length 
$L$ and sophisticated numerical methods shall be introduced to mitigate this unfavorable scaling~\cite{JensPEPSarehard,PhysRevB.99.195153,PhysRevB.90.064425, PhysRevE.102.032130, doi:10.1143/JPSJ.65.891}. On the contrary, the MERA in 2D is able to calculate expectation values exactly while satisfying area law but suffers from an even higher algorithmical complexity (at least $\mathcal{O}\left(m^{16}\right)$)~\cite{Eisert_2010}. 
Another well established approach is to extend the MPS for 2D systems~\cite{Stoudenmire_2012}: this approach has a very low algorithmic complexity ($\mathcal{O}\left(m^{3}\right)$), however, it is limited by the exponential scaling with the second dimension. 
As a compromise, TTNs are equally scalable in both system dimensions while still benefiting from a low numerical complexity, $\mathcal{O}\left(m^{4}\right)$ 
 but fail to capture area law for large systems sizes~\cite{FerrisTTNAreaLaw2013,TTNA19}. 

Hereafter, we introduce a novel ansatz {which augments the TTN} and show that is able to encode the area law keeping constant the algorithmic complexity to ($\mathcal{O}(m^4)$). Thus, the augmented Tree Tensor Network (aTTN) allows to tackle
open challenges in two- and three dimensional systems at sizes inaccessible before.

\paragraph{{Augmented Tree Tensor Network --}}\label{sec:aTTTN}
The aTTN ansatz
{$
|\psi_\text{aTTN}\rangle = \DLayerd |\psi_\text{TTN}\rangle
$}
is based on a TTN wave-function (a binary tree) $|\psi_\text{TTN}\rangle \in \otimes_i^N \Hilb_i$, with $\Hilb_i = \mathbb{C}^d$, with an additional sparse layer $\DLayer = \prod_k u_k$ of two-site unitary operators $\{u_k\}$ acting on (some of) the physical links of the TTN (see Fig.~1).
The additional layer $\DLayer$ contains $N_D$ independent non-overlapping 
 (i.e., acting on different couples of sites of the lattice $\Lat$) and thus commuting disentanglers $\{u_k\}$. 
In this way, $\DLayer$ describes a unitary mapping of the Hamiltonian $\Ha$ to an auxiliary Hamiltonian $\Ha_\text{aux} = \DLayer\Ha\DLayerd$. 
Each local transformation $u_k$ aims to decouple - or \textit{disentangle} in the spirit of the MERA language~\cite{MERA_bible} - entangled degrees of freedom in the quantum many-body state, that are then trivially included in the TTN layer. 
As described in the following, $\DLayer$ modifies the TTN in such a way that the aTTN satisfies the area law while keeping the complexity for the optimisation at $\mathcal{O}\left(m^4 \right)$. Thus, the aTTN 
overcomes the drawback of the TTN while maintaining its main advantages: {\it (i)} the low scaling with the bond dimension $m$ compared to both MERA and PEPS, and {\it (ii)} the ability to contract the network exactly. We stress that the aTTN  can be applied straighforwardly to a general $D$-dimensional system. Finally, we notice that the aTTN is effectively a particular subclass of a MERA, where the structure scale-invariance is traded for efficiency, as the scale-invariance is not necessary to ensure the area law at the tensor structure level.   
Fig.~\ref{fig:intro}a reports
an illustrative example of an aTTN for a two-dimensional $8\times 8$ system with the  $\DLayer$ layer composed by $6$  disentanglers $u_k$ (green). Notice that, 
not every physical site $j$ is addressed by a disentangler, a key property for preserving numerical efficiency. 
Indeed, the disentanglers positions is critical in order to {\it (i)} keep an optimal numerical complexity for the optimisation and {\it (ii)} efficiently encode an area law in the TN.

\paragraph{{Area Law in aTTN ---}}
 Hereafter, we specialize the discussion for the case of a two-dimensional square lattice $\Lat$ with $N=L\times L$ sites, and $L=2^n$. Moreover, we consider a binary TTN, where the tree tensors coarse-grain neighboring sites for each layer $\Lambda_l$ alternatingly along the $x$- (for even $l$) and the $y$-direction (odd $l$) with $l$ going from $l=1$ addressing the topmost layer to $l=\log{L}$ for the lowest layer (Fig.~\ref{fig:intro}a
). 
Each link $\nu$ of the tree bipartites the whole system $\Lat$ into two subsystems $\mathcal{A}^{[\nu]}$ and $\mathcal{B}^{[\nu]}$, separated by the boundary $\partial_\nu$ with length $\gamma_\nu$.
The area law implies that the entanglement entropy of the bipartition $S(\mathcal{A}^{[\nu]})$ (or $\mathcal{B}^{[\nu]}$ respectively) scales with $\gamma_\nu$. Thus, in order to faithfully represent the area law, the bond dimension $m_\nu$ of each link $\nu$ should scale with
$
m_\nu \approx e^{c \gamma_\nu}
%\label{eq:bd_area_}
$
where $c$ is a constant factor. This scaling argument implies that for 2D the TTN ansatz requires an exponentially large bond dimension $m$ within the topmost layers, for which $\gamma_\nu\sim L$. In conclusion, with increasing $L$
a TTN representation eventually fails to capture area law states
properties as it becomes exponentially inefficient. 
This necessary exponential scaling of the bond dimension can be prevented by inserting the tensors layer $\DLayer$ that augments the TTN with $N_D=\sum_\nu K_\nu$ disentanglers, where $K_\nu$ is the number of disentanglers along the boundary $\partial_\nu$ for each link $\nu$.
More precisely, each disentangler $u_k$ is positioned such that it acts on one physical site in the subsystems $\mathcal{A}^{[\nu]}$ {and the other in subsystems $\mathcal{B}^{[\nu]}$. Thus, each disentangler can maximally asses information in a $d^2$-dimensional space belonging to two local Hilbert spaces, reducing the entanglement for the TTN up to the order of $d^2$. As a result, all the $K_\nu$ disentanglers support the TTN link $\nu$ by disentangling information in the order of
$
   m_{\nu,\text{aux}} \approx (d^2)^{K_\nu} ~ .
$
Therefore, when applying $\DLayer$, the information assessed by the aTTN for the bipartition defined by each link $\nu$ scales with
$
m_{\nu, \text{eff}} \approx m_{\nu,\text{aux}} m_\nu = d^{2K_\nu + \xi_\nu }
$
, where we introduced the parameter $\xi_\nu \equiv \log_d{m_\nu}$ describing the contribution of the TTN bond dimension $m_\nu$. 
If we now impose $K_\nu \sim \gamma_\nu$, we obtain the exponential scaling required to encode the area law for the two-dimensional aTTN state.
Notice that the number of disentanglers $\Gamma_l=\sum_{\nu \in \Lambda_l} K_\nu$ for each layer shall be directly proportional to $
\sum_{\nu\in \Lambda_l} {\gamma_\nu} \sim L$.
 However, placing exactly $L$ disentanglers for each layer of the tree may lead to an unfavorable, $L$-dependent scaling of $\mathcal{O}(m^4d^L)$ for the computational complexity. Thus, a careful balance between the position of the disentanglers and their density has to be found.
This balance can be found as 
when no couple of disentanglers is directly connected by an Hamiltonian interaction term,
the algorithmic scaling remains of the order $\mathcal{O}(m^4d^2)$ (see the SM for details). Moreover,  the area law is still satisfied 
 removing the disentanglers crossing the boundaries of the bipartitions $\partial_\nu$ corresponding to the lower layers of the tree ($l\rightarrow \log{L}$). On the contrary, one shall keep the maximal allowed number of disentanglers (i.e., not connected by Hamiltonian terms) to support the boundaries corresponding to the higher branches ($l\rightarrow 1$).  
 Indeed, for $\nu\in \Lambda_l$ with $l\rightarrow \log{L}$,
the TTN bond-dimension $m_\nu$
is sufficiently large to capture the area law entanglement - or even the complete state - accurately, especially for reasonably small local dimensions $d$. Instead, the contribution $\xi_\nu$ of the TTN is negligibly small for $\nu \in \Lambda_l$ with $l \rightarrow 1$ compared to the required exponentially large bond-dimension, calling for the support of the disentanglers. 
In conclusion, as we numerically confirm hereafter, it is possible to engineer the disentangler positions in $\DLayer$, keeping both computational efficiency and the area-law fulfilled resulting in high-precision results also for large system sizes.

\paragraph*{{Ising model ---}}
We first benchmark the aTTN ansatz via a ground state search on the 2D Ising model with periodic boundary conditions. 
We consider a $L\times L$ lattice with 
$L=\{8,16,32,64\}$ and the Ising Hamiltonian
$
\mathcal{H}=\sum_{i,j=1}^L {\sigma^x_{i,j} \sigma^x_{i+1,j} + \sigma^x_{i,j} \sigma^x_{i,j+1} } + \sum_{i,j=1}^L \sigma^y_{i,j}  ~ .
$
$\sigma^\gamma_{i,j}$ (with $\gamma \in \{x, y, z\}$) are Pauli matrices acting on the site $(i,j)$. 
For small system sizes ($L=8$ and $L=16$) both the TTN and the aTTN reach the chosen machine precision of \texttt{1E-8} with high bond dimension. However, as expected, for larger sizes we find a significant improvement in the precision of the aTTN simulations.
Indeed, the different performances become evident for $L=32$ and $L=64$, as the aTTN and the TTN converge with increasing bond dimension to different values for the energy
Fig.~\ref{fig:intro} reports 
the relative error $\epsilon_m = |(\langle \mathcal{H}\rangle_m - E_{ex})/E_{ex} |$ for increasing bond dimension $m$ with respect to the energy $E_{ex}$ obtained by extrapolating the results of the aTTN for $L=8$ and $L=64$ (For the $L=16,32$ results see Fig.~\ref{fig:Ising} in the Supplementary material).
%%%%%% CORRELATIONS AND ENTROPY %%%%%%%%%%%%

\begin{figure}[t!]
\centering
  \includegraphics[width=0.85\linewidth]{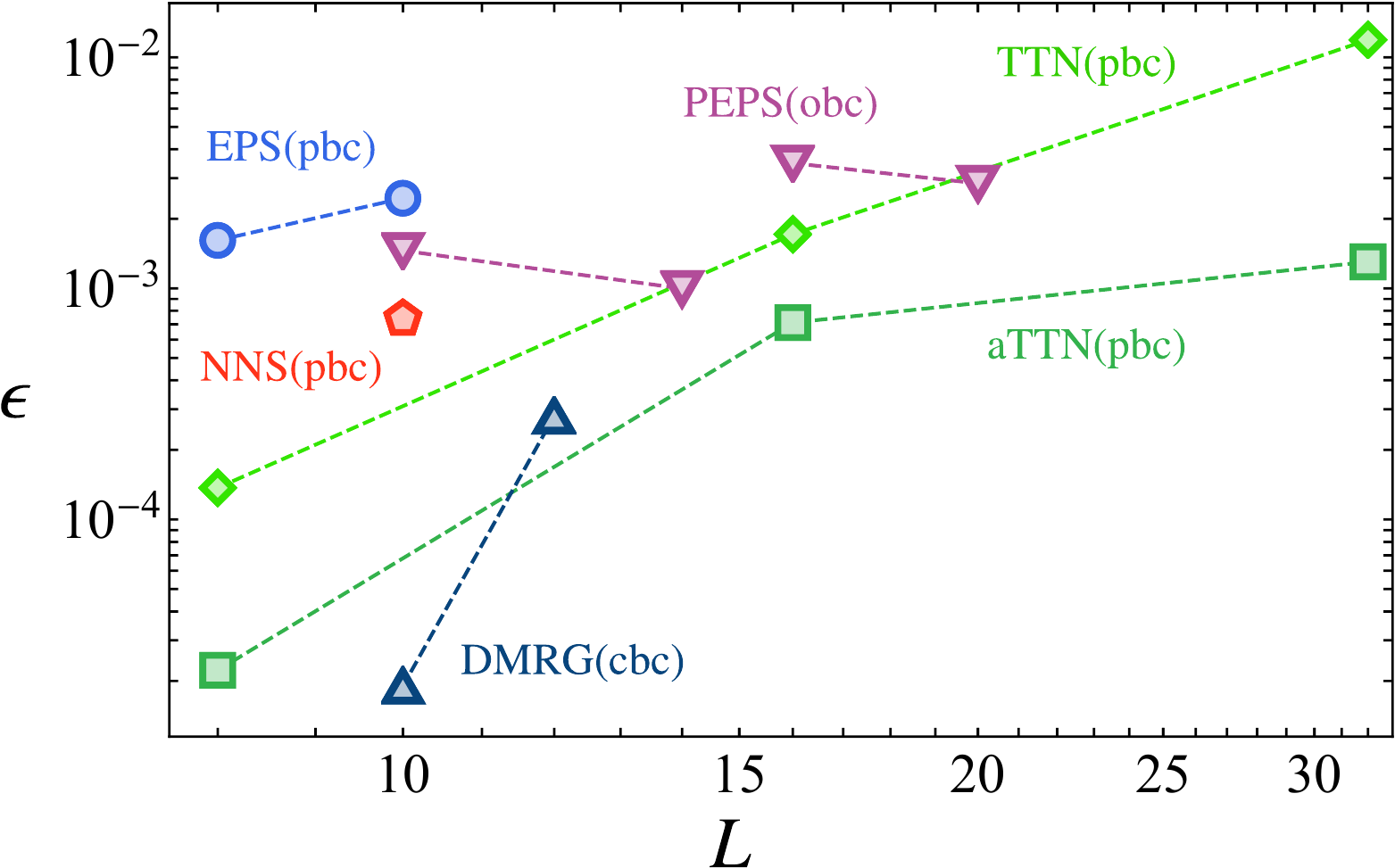}
  \captionsetup{justification=centerlast}
\caption{\label{fig:Heis} Relative error $ \epsilon$ of the 2D Heisenberg ground-state energy as a funciotn of the system linear size $L$ compared with the best
available estimates obtained by MC~\cite{Sandvik_1997} for the TTN, aTTN, NNS~\cite{Carleo_2017}, EPS~\cite{Mezzacapo_2009}, PEPS~\cite{Lubasch_2014}, 2D-DMRG~\cite{Stoudenmire_2012}. Depending on the method open (obc), cylindrical (cbc) or periodic (pbc) boundary conditions have been chosen. For each datapoint, we compare the Monte Carlo result with the same boundary conditions.}
\end{figure}
%%%%%%%%%%%%%%%%%%%%%%%%%%%%%%%%%%%%%%%%%%%%

\paragraph*{{Heisenberg model ---}} We now analize the more challenging critical antiferromagnetic two dimensional Heisenberg model 
$
\mathcal{H}=\sum_{i,j=1}^L \sum_{\gamma\in \{x, y, z\}} { \sigma^\gamma_{i,j} \sigma^\gamma_{i+1,j} + \sigma^\gamma_{i,j} \sigma^\gamma_{i,j+1} },
$
with periodic boundary conditions. 
In Fig.~\ref{fig:Heis} we compare the estimated energy density obtained by extrapolating the results from the TTN and the aTTN at $m\rightarrow \infty$ with previous results from different variational ans\"atze. In particular, we plot the relative error obtained by the different tensor network ans\"atze and the best known results, obtained via Quantum Monte Carlo~\cite{Sandvik_1997}.
Differently from the Ising model, we find the aTTN to be more accurate than the TTN even at lower system sizes, such as $L=8,16$. Interestingly, the aTTN for $L=16$ obtains an even more precise ground state energy density compared to most of the alternative variational ans\"atze at lower finite system size of $L=10$, such as Neural Network states, Entangled Pair States or PEPS~\cite{Carleo_2017,Mezzacapo_2009,Lubasch_2014}. We mention that, while the PEPS is very efficient with its ability to work directly in the thermodynamic limit in describing infinite systems as iPEPS~\cite{Jordan_2008,PhysRevB.80.094403}, the PEPS analysis for finite sizes are, for now, limited to $N=20\times 20$ systems. It turns out that for this model a very competitive variational approach is the 2D-DMRG, which outperforms the alternative methods for finite sizes with open or cylindrical boundary conditions up to the system size $L=12$, but struggles with periodic boundary conditions and with increasing both system sizes $L \gtrsim 12$~\cite{Stoudenmire_2012}.
Finally, we extended our analysis to reach the system size of $L=32$ for which, to the best of our knowledge, no public result is available. Thus, we estimated the error, extrapolating the value of the finite size scaling of Monte Carlo~\cite{Sandvik_1997}.

We point out that the here performed aTTN simulations (as well as the TTN simulations) exploit a $U(1)$ symmetry. However, for this model, we could further drastically improve the performance of the aTTN by incorporating the present $SU(2)$ symmetry in the simulation framework~\cite{TTNA19, sv13, singh2012tensor}.

\paragraph*{{Interacting Rydberg atoms ---}}

We now present new physical results, on a long-range interacting system by studying the zero-temperature phase diagram of an interacting Rydberg atoms two-dimensional lattice~\cite{Bernien2017}, described by the Hamiltonian
$
\mathcal{H}_{ryd}=\sum_{\vec{r}}[\frac{\Omega}{2}\sigma_{\vec{r}}^x\,
-\Delta n_{\vec{r}} + \frac{1}{2}
\sum_\vec{s}V(|\vec{r}-\vec{s}|) n_{\vec{r}}  n_\vec{s}
]
$
where the Rabi frequency $\Omega$ couples the ground $\ket{g}_{\vec{r}}$ and the excited Ryderg state $\ket{r}_{\vec{r}}$
and $n_{\vec{r}}=\ket{r}\bra{r}_{\vec{r}}$. $\Delta$ is the detuning and $V(|\vec{r}-\vec{s}|) = c_6/|\vec{r}-\vec{s}|^6$ is the interaction strength between two excited atoms placed at sites $\vec{r}$ and $\vec{s}$. We keep the interaction terms up to the fourth-nearest neighbor and 
set the Rabi frequency $\Omega=4 \,\mathrm{M\,Hz}$, while the interaction parameters refer to $^{87}\mathrm{Rb}$ atoms excited to the state $\ket{70 S_{1/2}}$, for which $c_6=863\,\mathrm{GHz}\, \mu m^6$.

The interactions limit the maximum excitation density according to  the Rydberg blockade radius $r^*$ -- the minimum distance at which two atoms can be simultaneously excited -- defined by the relation $V(r^*)=\Omega$. 
The competition between the interactions strength and $\Delta$ generates non-trivial phases characterized by regular spatial excitation-density distributions. The Fig.~\ref{fig:phase_diagr}a shows the phase diagram of the system as a function of the detuning and the nearest-neighbor interaction energy $V_{nn}$, obtained via aTTN simulations with $L=4,8,16,32$ with open boundary conditions. 

For low values of the detuning $\Delta$, the system exhibits a disordered phase characterized by the absence of excitations while, increasing $\Delta$, excitations are energetically favored and the interactions determine their spatial arrangement. In the limit of $V_{nn}\rightarrow 0$, or $a\rightarrow \infty$, the atoms are non-interacting and the expectation value $\langle n_{\vec{r}}\rangle\rightarrow  1$ for $\Delta\ll \Omega$. 
At larger values of $V_{nn}$, corresponding to $r^*/\sqrt{2}<a<r^*$, nearest neighbor atoms cannot be simultaneously excited, giving rise to the $\mathbb{Z}_2$ phase \cite{Bernien2017,Schau1455} with a two-degenerate ground state with the excitations distributed in a chess-bond like configuration, as shown in Fig.~\ref{fig:phase_diagr}a. Nevertheless, the $\mathbb{Z}_2$ disappears at low values of $V_{nn}$ and large detuning, as all the atoms are excited (light orange, right-bottom region of the phase diagram). 
\begin{figure}
   %\hspace{2cm}
    \centering
    \begin{tabular}{c}
 \includegraphics[width=0.9\linewidth]{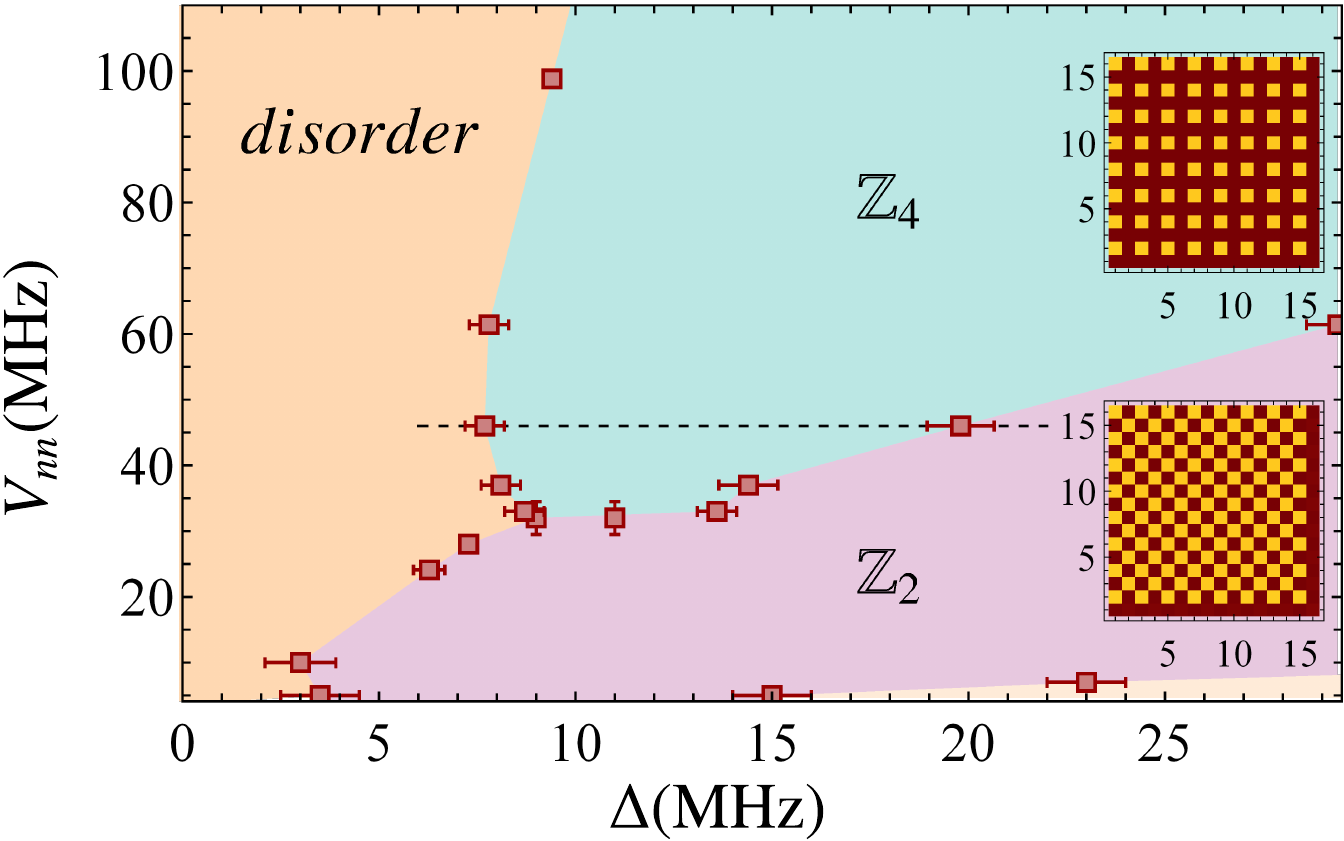} \\
\\
     \includegraphics[width=1\linewidth]{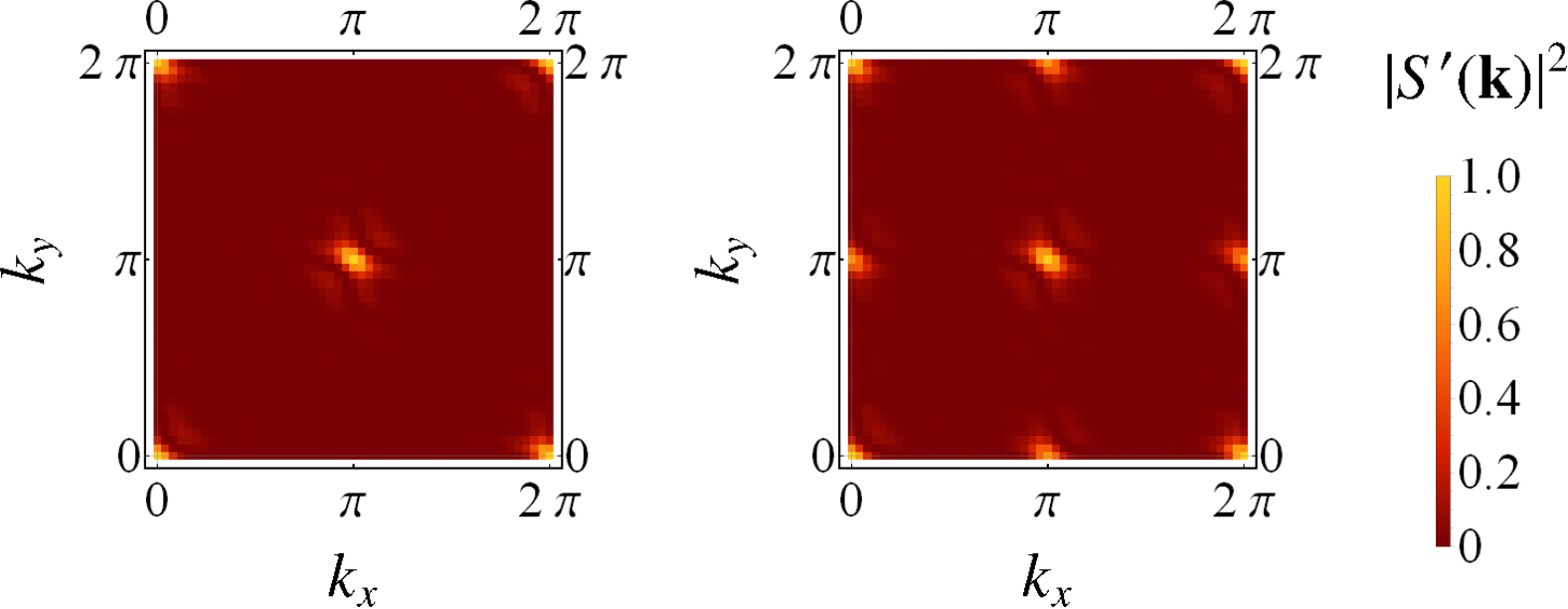} 
      \end{tabular}
   \\
   \pgfputat{\pgfxy(0.4,8.5)}{\pgfbox[left,bottom]{\footnotesize (a)}} 
     \pgfputat{\pgfxy(0.0,3.8)}{\pgfbox[left,bottom]{\footnotesize (b)}}       \pgfputat{\pgfxy(3.9,3.8)}{\pgfbox[left,bottom]{\footnotesize (c)}}
 \captionsetup{justification=centerlast}
 \caption{\label{fig:phase_diagr}Up: Phase diagram as a function of the detuning $\Delta$ and the nearest-neighbors interaction energy $V_{nn}$. 
 The disordered phase is characterized by a substantially uniform distribution of the excitations, while in the phases $\mathbb{Z}_2$ and $\mathbb{Z}_4$ the excitations are distributed as shown in the upper ($\mathbb{Z}_4$) and lower ($\mathbb{Z}_2$) insets. 
 Down: Renormalized structure factor $S'(\vec{k})=S(\vec{k})/S(\vec{0})$ for $V_{nn}=46\,\mathrm{MHz}$ and (a) $\Delta=28\,\mathrm{MHz}$ ($\mathbb{Z}_2$ phase) and (b) $\Delta=12\,\mathrm{MHz}$ ($\mathbb{Z}_4$ phase). 
 Other parameters: $\Omega=4 \,\mathrm{MHz}$.
 }
\end{figure} 
The spatial distribution of the excitations in the orderded phase is well captured by the peaks of the static structure factor
$
S(\vec{k})=\frac{1}{N^2}\sum_{\vec{r},\vec{s}}\nep^{-i\vec{k}\cdot(\vec{r}-\vec{s})}\langle n_\vec{r}  n_\vec{s}\rangle\,.
$
In particular, the phase $\mathbb{Z}_2$ exhibits a peak in $(\pi,\pi)$, as shown in Fig. \ref{fig:phase_diagr}b. 
The transition from the disordered to the $\mathbb{Z}_2$ phase is a second-order one, as it emerges by computing the second derivative of the energy with respect to $\Delta$ (see Supplementary Material).
In order to determine the critical line separating the two phases we define the nonlocal order parameter $O^{(2)}_{\vec{r}} =(n_{r_x,r_y} - n_{r_x+1,r_y}- n_{r_x,r_y+1}+ n_{r_x+1,r_y+1})/4$ and perform a finite-size scaling analysis of   $\langle O^{(2)\dagger}_\vec{r}  O^{(2)}_\vec{r}\rangle$ \textit{vs} $\Delta$, where $\langle O^{(2)\dagger}_\vec{r}  O^{(2)}_\vec{r}\rangle$ is estimated by $ S({\pi,\pi})$~\cite{Binder1981, PhysRevB.99.094434} (see SM). 
By further reducing $a$, the blockade radius prevents diagonal-adjacent atoms to be excited. As a consequence, 
each one of the $\mathbb{Z}_2$ ground states breaks into two different states, giving rise to the four-degenerate phase $\mathbb{Z}_4$: In each one of the ground states of this phase, each excited atom is surrounded by atoms in their ground states (see upper inset in Fig.~\ref{fig:phase_diagr}a). We observe a second-order phase transition in $V_{nn}$ for $\Delta\simeq 10\,\mathrm{MHz}$ from the $\mathbb{Z}_2$ to the $\mathbb{Z}_4$ phase  at $V_{nn}^c=32\pm2.5\,\mathrm{MHz}$ (or equivalently $a=r^*/\sqrt{2}$).
The static structure factor exhibits four additional peaks in the points such as $(0,\pi)$ as shown in Fig.~\ref{fig:phase_diagr}c. 
As in the $\mathbb{Z}_2$ case, a second-order phase transition occurs between the disordered phase to the $\mathbb{Z}_4$ by changing $\Delta$ at a fixed $V_{nn}$. We determine the critical line by introducing 
the order parameter $O^{(4)}_\vec{r} =(n_{r_x,r_y} + i\, n_{r_x+1,r_y}-i\, n_{r_x,r_y+1}- n_{r_x+1,r_y+1})/4$, defined such that the value of $\langle O^{(4)\dagger}_\vec{r}  O^{(4)}_\vec{r}\rangle$ equals $S({0,\pi})$ in the $\mathbb{Z}_4$ phase.
Remarkably, we find that another second-order phase transition occurs by further increasing $\Delta$, leading the system from the $\mathbb{Z}_4$ to the $\mathbb{Z}_2$ phase. 
We expect that at larger values of $V_{nn}$ new phases would emerge and accordingly, new phase transitions would occur by changing $\Delta$.

\paragraph*{Conclusions ---}
We have augmented the well-established TTN geometry with
a new ansatz which reproduces area law for high dimensional quantum many-body systems. The aTTNs allowed us to reach unprecedented sizes ($32\times 32$) in the study of critical system, going beyond the {current} possibilities of PEPS and DMRG, and therefore set new benchmarks for future numerical simulations. 
As a first application of aTTNs, we have characterized the phase diagram of two dimensional Rydberg atoms in optical tweezers, with atoms number of the order of current and near future experiments~\cite{Browaeys2020}.
In conclusion, the aTTN ansatz introduced here provides a novel powerful tool for simulating quantum systems in two or higher-dimensions, which, beyond many interesting physical applications will provide benchmark near-future quantum simulations and computations on different platforms, as we have demonstrated for Rydberg atoms in optical tweezers.

\paragraph*{{Acknowledgements ---}}
This work is partially supported by the Italian PRIN 2017, INFN through the project ``QUANTUM” , Fondazione CARIPARO, the Horizon 2020 research and innovation programme under grant agreement No 817482 (Quantum Flagship - PASQuanS), the QuantERA projects QTFLAG and QuantHEP, and the DFG project TWITTER. We acknowledge computational resources by CINECA through the project IsC78 - EPM2D3, the Cloud Veneto, the BwUniCluster and by ATOS Bull. 

\bibliographystyle{apsrev4-1}

%\bibliography{references}


\clearpage

\appendix

\section{SUPPLEMENTARY MATERIAL}

\subsection{Optimisation aTTN}

To target the ground-state of a given Hamiltonian $\Ha$, we optimise the variational parameters of the aTTN wavefunction $\psi$ minimising the system energy
\be
E = \langle \psi | \Ha | \psi \rangle ~.
\ee

The complete optimisation procedure for the aTTN herefore consists of three different parts:
{\it (i)} The optimisation of the disentanglers $u_k$, {\it (ii)} the mapping of the Hamiltonian $\Ha\stackrel{\DLayer}{\longrightarrow} \Ha_\text{aux}$, and {\it (iii)} the optimisation of the internal TTN with the auxiliary Hamiltonian $\Ha_\text{aux}$.

In the following, we provide a brief description regarding each of the optimisation parts, while we refer to Ref.~\cite{aTTN_coming} for a more comprehensive technical description. Further, hereinafter, we assume the Hamiltonian $\Ha = \sum_p \Ha_p$ to be a sum of local interactions $\Ha_p$.

 %%%%%% SKETCH OF THE OPTIMISATION OF ONE DISENTANGLER %%%%%%%%%%%%
\paragraph{\textbf{ Disentangler Layer $\DLayer$ ---}}

The layer $\DLayer$ is optimised by addressing all the disentanglers $u_k$ one by one.
%optimising all of the disentanglers $u_k$ within $\DLayer$ individually. 
The fundamental procedure therefore origins from the general MERA optimisation~\cite{MERA_bible}. Accordingly, the energy $E = \langle \Psi | \Ha | \Psi \rangle$ depends bilinearly on one single disentangler $u_k$ and its complex conjugate $u_k^\dagger$,

\be
E(u_k) = \tr\left\lbrace \sum_{p} u_k M_p u_k^\dagger N_p \right\rbrace  + c_k ~,
\label{eq:en_dis}
\ee

where $M_p$ and $N_p$ reflect the contractions of the Hamiltonian part $\Ha_p$ with the environment around $u_k$ and $u_k^\dagger$. Further, $p$ only runs over the Hamiltonian parts $\Ha_p$ which act on one or both sites attached to the disentangler $u_k$. All the other Hamiltonian parts are included in the constant $c_k$ and independent of $u_k$. As described in detail in Ref.~\cite{MERA_bible}, for the optimisation of the disentangler $u_k$, we
\begin{itemize}
\item[(i)] compute the environment $\Gamma_k = M_p u_k^\dagger N_p $ of $u_k$ for attached all Hamiltonian parts $\Ha_p$

\item[(ii)] perform a singular value decomposition $ \Gamma_k = U\sigma V^\dagger$

\item[(iii)] update the disentangler $u_k \rightarrow u_k' = -VU^\dagger$

\item[(iv)] restart from (i) until all singular values $\sigma_i$ converge.
\end{itemize}

Once the disentangler $u_k$ is optimised we move on to the next disentangler, until the complete set of disentanglers within the layer $\DLayer$ have been optimised. 

We position the disentanglers so that two, or more, disentanglers do not share the same interaction part $\Ha_p$. In this way, the optimisation of each disentangler $u_k$ is completely independent of all the other disentanglers $u_{k'}$. Thus, it is sufficient to optimise each disentangler just once to obtain the optimised disentangler layer $\DLayer$ - and furthermore, all optimisations for each disentangler can be fully parallalised.

Applying this MERA optimisation scheme to the engineered aTTN architecture leads to a numerical complexity of $\mathcal{O}(m^4 d^2 + m^3d^4 + d^6)$. 

\paragraph{\textbf{ Hamiltonian mapping ---}}

Having optimised the disentangler layer $\DLayer$, we move on to map the physical Hamiltonian $\Ha$. In particular, the new layer of disentanglers $\DLayer$ maps $\Ha$ to an auxiliary Hamiltonian $\Ha_\text{aux} \equiv \DLayer \Ha \DLayerd$ as preconditioning for the optimisation of the TTN. The energy expectation value

\be
\langle \psi_\text{aTTN} | \Ha | \psi_\text{aTTN} \rangle = \langle \psi_\text{TTN} | \underbrace{\DLayer \Ha \DLayerd}_{\equiv \Ha_\text{aux}} | \psi_\text{TTN} \rangle
\label{eq:mapping}
\ee

equals the expectation value of the mapped Hamiltonian for the internal TTN wave-function. This transformation of the complete Hamiltonian $\Ha \stackrel{\DLayer}{\longrightarrow} \Ha_\text{aux}$ is done by contracting each interaction part $\Ha_p$ separately with $\DLayer$

\be
\Ha_{p,\text{aux}} = \DLayer \Ha_p \DLayerd = \prod_k{u_k} \Ha_p \prod_{k'} {u_{k'}^\dagger}
\ee

We recall that we permit more then one disentangler to be applied on the subspace of a single Hamiltonian part $\Ha_p$ and that all disentanglers strictly obey the isometry condition $u_k u_k^\dagger = \Id$, implying $\DLayer \DLayerd = \Id$. Thus, the mapping of each interaction $\Ha_p$ either {\it (i)} becomes trivial $\Ha_{p,\text{aux}}\rightarrow\Ha_p$ if $\Ha_p$ acts on physical sites where no disentanglers are placed, as in this case $\Ha_p$ commutes with $\DLayer$, or {\it (ii)} equals $\Ha_{p,\text{aux}}\rightarrow u_{k^p}\Ha_pu_{k^p}^\dagger$ where $u_{k^p}$ is the disentangler acting on the subspace addresed by $\Ha_p$.

At this point, we confirm that, in general, it is possible to lift the above-mentioned restriction on the positioning of the disentanglers, such that two or more disentanglers may indeed be applied on the same Hamiltonian part $\Ha_p$. However, in this case, the resulting operator after the mapping $\Ha_{p,\text{aux}}$ can become highly non-trivial and may even contain several loops in its structure, like e.g. a PEPO~\cite{PhysRevB.92.035152}. 
This might in the worst case increase the numerical complexity for the optimisation of the internal TTN, and thereby the total numerical complexity for the aTTN, to a with system size exponential scaling $\mathcal{O}(m^4d^L)$ dependency. Thus, when constraining the aTTN in its construction to prevent an interaction part $\Ha_p$ being addressed by two - or more - disentanglers, we not only simplify the technical implementation and enable the possibility to parallelise the optimisations of each disentanglers but further restrict the numerical cost to a worst-case scaling of $\mathcal{O}(m^4d^2)$. 

\paragraph{\textbf{TTN optimisation ---}}
After mapping the Hamiltonian, we optimise the internal TTN following the well-established optimisation procedure prescriptions of Ref.~\cite{TTNA19}. 
Thus, we sweep through the TTN structure from the bottom to the top, solving the underlying eigenvalue problem locally for each tensor in order to optimise the complete TTN structure globaly. More precise, we iteratively optimise each pair of tensors separately exploiting the subspace-expansion technique which approximates a complete two-site update for each local optimisation. Thereby, we are able to optimise the complete internal TTN structure with a numerical complexity of $\mathcal{O}\left( m^4 \right)$ (compared to $\mathcal{O}\left( m^6 \right)$ for the two-site optimization). For each optimization of a local tensor, we exploit the Arnoldi algorithm implemented in the ARPACK library to solve the underlying eigenvalue problem for the single tensors.

\paragraph{\textbf{Further practical remarks ---}}

We begin each simulation by performing a few iterations on the internal TTN without including the disentangler layer $\DLayer$. Therefore, we randomly initialise the TTN. In the first optimisation steps, we set a low precision in the Arnoldi algorithm for solving the local eingenvalue problems and use a reasonably large space-expansion in order to derive at a TTN which is able to give a reasonable qualitative representation of the ground state. Consequently, we initialise the disentangler layer $\DLayer$ as a set of identities $\{u_k\}\equiv \{\Id\}$ positioned at the sites we aim to disentangle. With the resulting aTTN, we iteratively perform the optimisation procedure described above: optimising $\DLayer$, mapping $\Ha$ to $\Ha_\text{aux}$ and optimising the internal TTN. We repeat the complete procedure until the energy $E = \langle \psi_\text{aTTN} | \Ha | \psi_\text{aTTN} \rangle$ convergences.

We further point out that we are able to exploit symmetries in the aTTN as well as in the TTN simulations. In particular, for the Ising model and the Rydberg study we incorporated the underlying $\mathcal{Z}_2$ symmetry, while for the Heisenberg model we exploited a $U(1)$ symmetry. For the latter, however, we could further drastically improve the performance of the aTTN by incorporating the present $SU(2)$ symmetry in the simulation framework~\cite{TTNA19, sv13, singh2012tensor}.

\subsection{Calculation of Observables}\label{sec:dressed}

In accordance with Eq.~(\ref{eq:mapping}), when calculating the expectation value of an observable $O$, we compute

\be
\langle \psi_\text{aTTN} | O | \psi_\text{aTTN} \rangle = \langle \psi_\text{TTN} | \underbrace{\DLayer O \DLayerd}_{\equiv O_\text{aux}} | \psi_\text{TTN} ~. \rangle
\ee

Thus, each observable is mapped by $\DLayer$ to $O_\text{aux}$, before we compute the expectation value of $O_\text{aux}$ for the internal TTN. The latter computation is done following the well-established receipt for calculating expectation values with TTNs~\cite{TTNA19, SimoneBook}. In the former, we follow the same idea as in the mapping of Hamiltonian parts discussed above.

In our analysis, we restricted ourselves to local observables (acting on a single physical site $i$ only) and two-body operators, or correlators (acting on two different physical sites $i$ and $j$). The mapping of the former (as the mapping of an interaction $\Ha_p$ in the prior section) either {\it (i)} becomes trivial $O_{\text{aux}}\rightarrow O_p$ if no disentanglers is placed at the local site $i$ or else {\it (ii)} equals $O_{\text{aux}}\rightarrow u_{k} O u_{k}^\dagger$ where $u_{k}$ is the disentangler acting on site $i$. In the mapping procedure of a two-body operator acting on sites $i$ and $j$, however, there can be a third scenario in which two different disentanglers $u_k$ and $u_k'$ act on the two different sites $i$ and $j$. In this case, we contract the operator with both disentanglers resulting in the auxiliary operator $O_{\text{aux}} = (u_{k}) (u_{k'}) O (u_{k}^\dagger) (u_{k'}^\dagger)$ for the TTN computation. After this mapping the computation equals a calculation of the expectation value of a string observable for the TTN, which exhibits a worst-case complexity of $\mathcal{O}(m^4 \log{N})$.

\subsection{Disentangler Positioning}\label{app:DEEngineering}

The positioning of the disentanglers is a crucial part of the successfully aTTN application. On the one hand, we want as many disentanglers as possible to enforce the internal TTN structure and indeed enable the aTTN to capture the underlying area law. On the other hand, we permit more then one disentangler to be applied on the subspace of a single Hamiltonian part $\Ha_p$ for the sake of computational benefits, namely {\it (i)} the possibility to parallelise the optimisations of the disentanglers, {\it (ii)} a simplification in the technical implementation and most importantly {\it (iii)} avoiding a with system size exponential scaling of $\mathcal{O}(m^4d^L)$ in the worst-case of the optimisation. Thus, this problem becomes a non-trivial engineering task.

 %%%%%% SKETCH OF THE MAPPING BY THE DISENTANGLER LAYER %%%%%%%%%%%%
\begin{figure*}[t!]
\begin{center}
\centering
\begin{minipage}{0.35\textwidth}    
  \centering
  \includegraphics[width=0.5\textwidth]{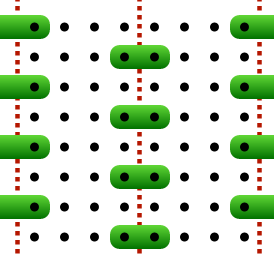} \\ (a) \\ \vspace{0.4cm}
  \includegraphics[width=0.9\textwidth]{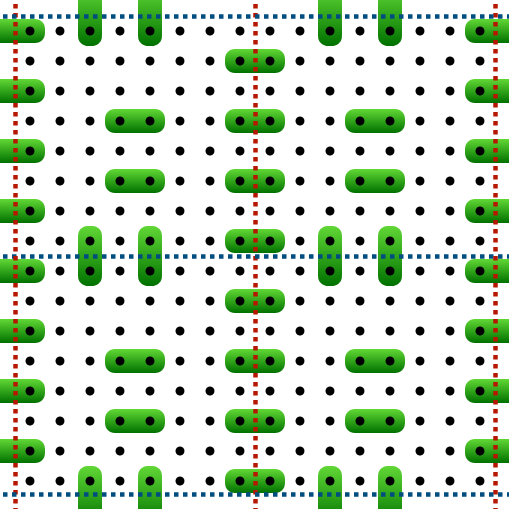} \\ (b)
\end{minipage}
\begin{minipage}{0.6\textwidth}
  \centering
  \includegraphics[width=0.9\textwidth]{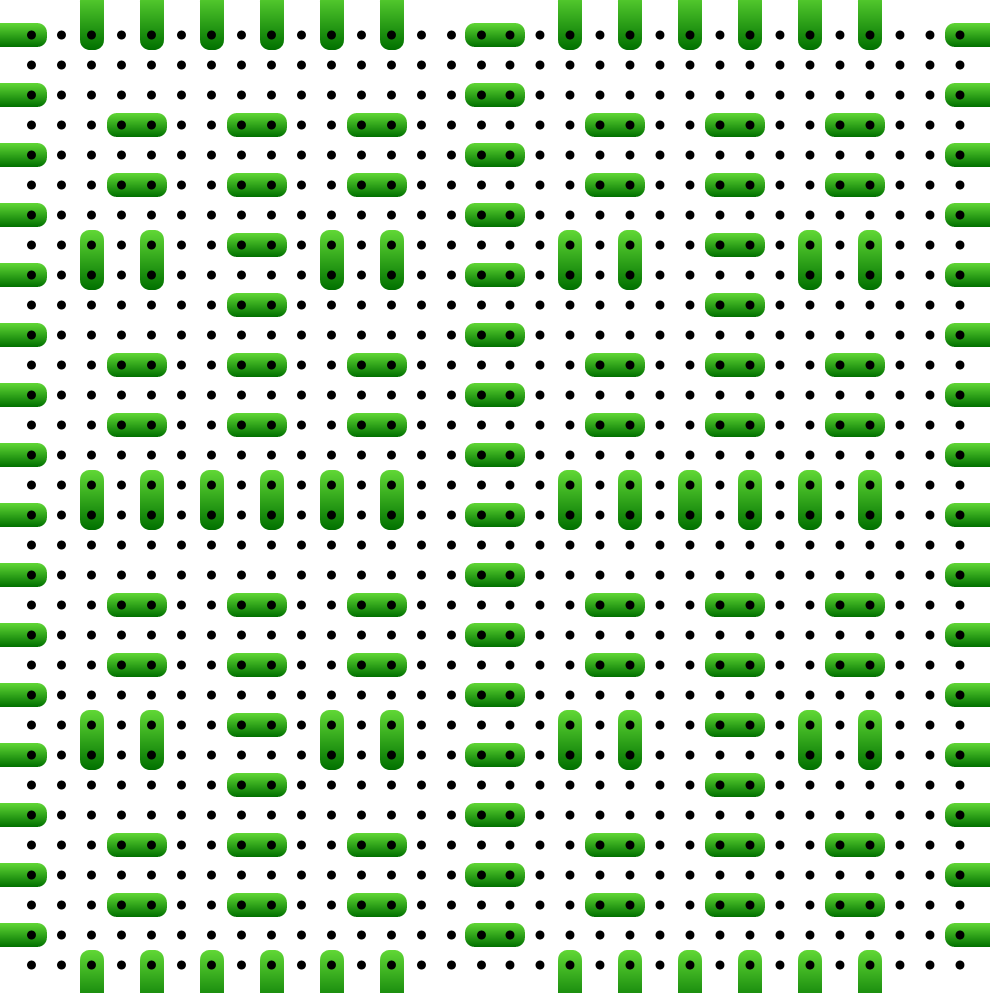} \\ (c)
\end{minipage}

\caption{\label{fig:DE_pos}
Disentangler positions for simulations of Hamiltonians with periodic boundary conditions and nearest-neighbor interactions only for system size $L=8$ (a), $L=16$ (b), $L=32$ (c). The red(blue)-dotted lines indicate the bipartition of the up-most(second highest) link of the internal TTN.
}
\end{center}
\end{figure*} 
%%%%%%%%%%%%%%%%%%%%%%%%%%%%%%%%%%%%%%%%%%%%%%%%

Recalling the notation introduced in the main text, $K_\nu$ is the number of disentanglers placed along the boundary $\partial_\nu$ of the link $\nu$ and $\Gamma_l=\sum_{\nu \in \Lambda_l}K_\nu$ the number of disentanglers associated to the links of the layer $\Lambda_l$.
Fig.~\ref{fig:DE_pos} shows the positioned disentanglers for our analysis regarding Hamiltonians with periodic boundary conditions and nearest-neighbor interactions only, i.e. the Ising model and Heisenberg model. The red-dotted lines indicate the bipartition of the up-most link of the internal TTN. In our strategy, we start placing as many disentanglers as possible to reinforce the top-most link, then place as many disentangler as possible to reinforce the second highest links (indicated in blue), and so on. Following this strategy, we position $N_D=\Gamma_1=8$ disentanglers for a $8\times 8$ system, all of which supporting the highest TTN layer ( see Fig.~\ref{fig:DE_pos} (a) ). In the $16\times 16$ analysis, we place $\Gamma_1=16$ supporting the highest, $\Gamma_2=8$ the second highest and $\Gamma_3=8$ the third highest TTN layer ( see Fig.~\ref{fig:DE_pos} (b) ). For the $32\times 32$ systems, we place in total $N_D=128$, with $\Gamma_1=32$, $\Gamma_2=24$, $\Gamma_3=24$, $\Gamma_4=16$ and $\Gamma_5=32$ ( see Fig.~\ref{fig:DE_pos} (b) ).

 %%%%%% SKETCH OF THE MAPPING BY THE DISENTANGLER LAYER %%%%%%%%%%%%
\begin{figure*}[t!]
\begin{center}
\centering
\begin{minipage}{0.35\textwidth}    
  \centering
  \includegraphics[width=0.5\textwidth]{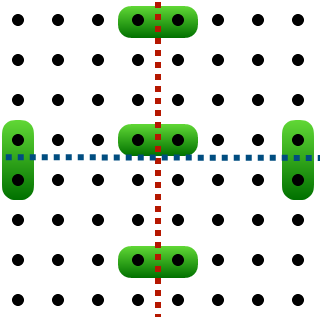} \\ (a) \\ \vspace{0.4cm}
  \includegraphics[width=0.9\textwidth]{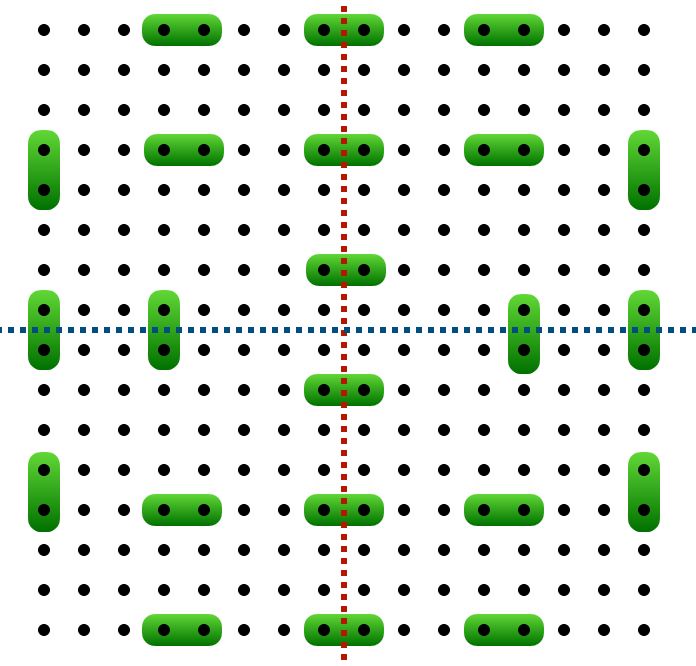} \\ (b)
\end{minipage}
\begin{minipage}{0.6\textwidth}
  \centering
  \includegraphics[width=0.9\textwidth]{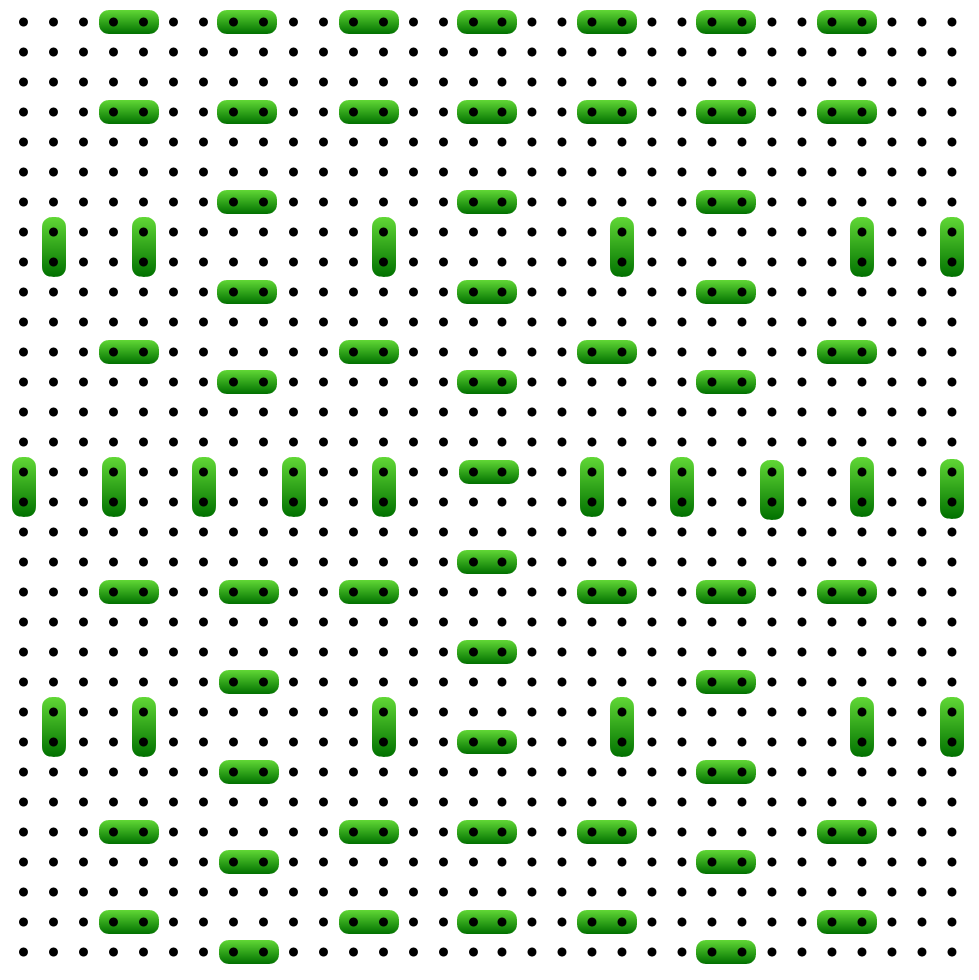} \\ (c)
\end{minipage}

\caption{\label{fig:DE_pos_Ryd}
Disentangler positions for system size $L=8$ (a), $L=16$ (b), $L=32$ (c) for the Rydberg simulations, i.e. for Hamiltonians with open boundary conditions and nearest-neighbor and next-nearest-neighbor interactions. The red(blue)-dotted lines indicate the bipartition of the up-most(second highest) link of the internal TTN.
}
\end{center}
\end{figure*} 
%%%%%%%%%%%%%%%%%%%%%%%%%%%%%%%%%%%%%%%%%%%%%%%%

In contrast to the analysis of the Ising model and Heisenberg model, the Rydberg computations deal with open boundary conditions and we consider next-nearest-neighbor terms in the Hamiltonian, i.e. with the lattice spacings (0,1), (1,0), (1,1), (1,-1), (2,0), (0,2), (2,1), (2,-1), (1,2), (1,-2). Therefore, the disentanglers are positioned more distant to each other with at least two empty sites between each other (compare Fig.~\ref{fig:DE_pos_Ryd}). Following the same strategy as before, we place as many disentanglers as possible, starting from the top-most link. Consequently, we position here $N_D=4$ (with $\Gamma_1=3$ and $\Gamma_2=2$) disentanglers for a $8\times 8$ system, $N_D=22$ (with $\Gamma_1=6$, $\Gamma_2=4$, $\Gamma_3=8$, $\Gamma_4=4$) for a $16\times 16$ system and $N_D=77$ (with $\Gamma_1=11$, $\Gamma_2=10$, $\Gamma_3=20$, $\Gamma_4=12$, $\Gamma_5=24$) for a $32\times 32$ system.

\subsection{Ising model}

%%%%%% CORRELATIONS AND ENTROPY %%%%%%%%%%%%

\begin{figure}[t!]
\centering
\begin{tabular}{cc}
\includegraphics[width=0.47\linewidth]{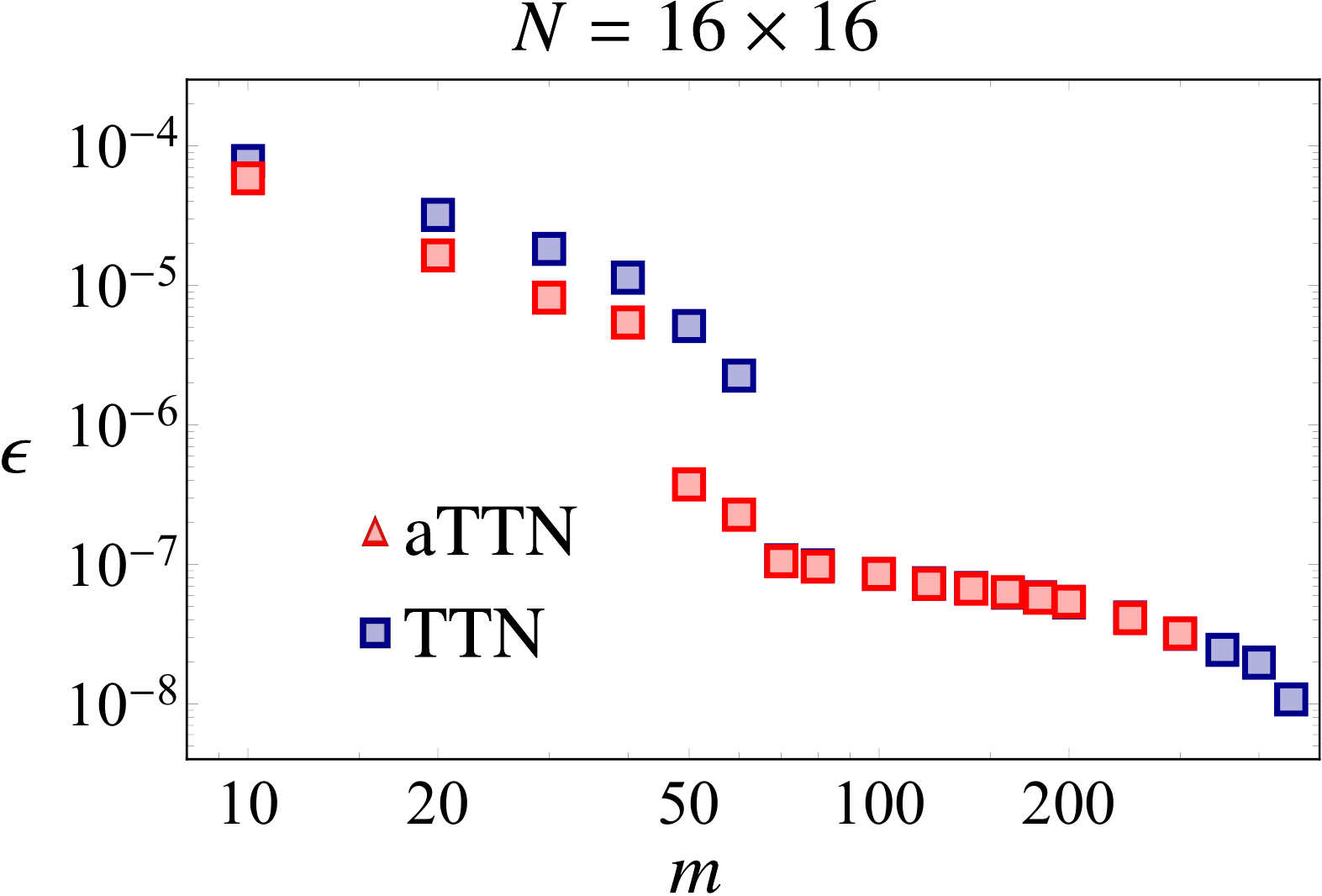} &
\includegraphics[width=0.47\linewidth]{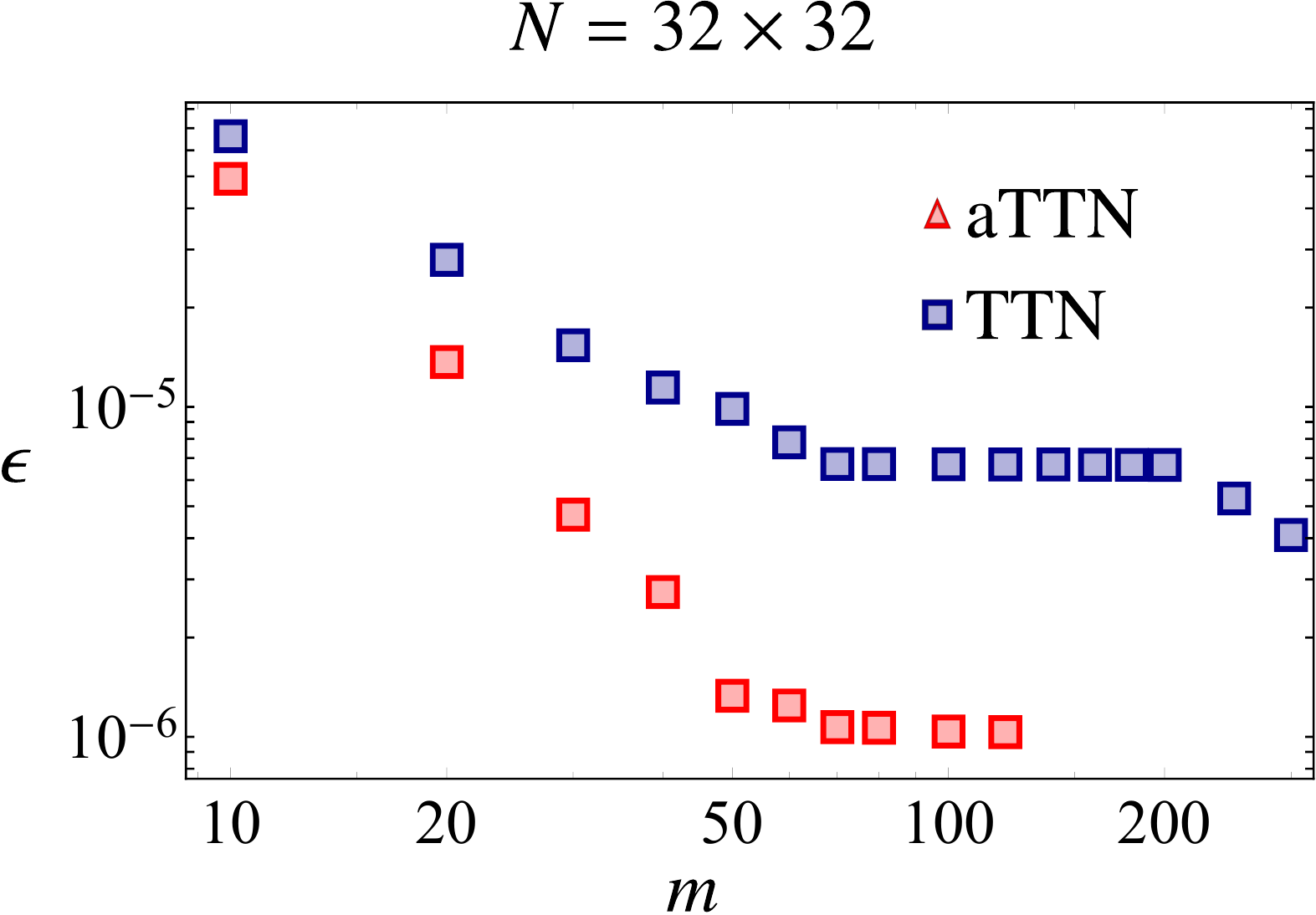}
\end{tabular}\\
\pgfputat{\pgfxy(0.5,-0.0)}{\pgfbox[left,bottom]{\footnotesize (a)}}       
\pgfputat{\pgfxy(4.5,0.35)}{\pgfbox[left,bottom]{\footnotesize (b)}}
\caption{\label{fig:Ising} Relative error $\epsilon$ of the 2D Ising ground-state energy compared with the best
available estimates obtained by extrapolating the results of the aTTN for the system sizes $L=\{16,32\}$.}
\end{figure}
%%%%%%%%%%%%%%%%%%%%%%%%%%%%%%%%%%%%%%%%%%%%

As referred to in the main text, we provide the comparison of the aTTN against the TTN for the ground state search on the 2D Ising model at system size  $L=\{16,32\}$. In Fig.~\ref{fig:Ising} we report the relative error $\epsilon_m = |(\langle \mathcal{H}\rangle_m - E_{ex})/E_{ex} |$ for increasing bond dimension $m$ with respect to the energy $E_{ex}$ obtained by extrapolating the results of the aTTN.
We point out that, for the lower system size of $L=16$, the TTN is as precise as the aTTN and the data points for both tensor networks analysis indeed overlap reaching the chosen machine precision of \texttt{1E-8} with high bond dimension. This confirms that the TTN is as much as the aTTN capable of capturing the entanglement in the 2D system for lower system size. However, going to larger system sizes, the TTN cannot hold up to the with $L$ exponentially growing area law and eventually fails to represent the ground state accurately, while the aTTN in contrast is able to maintain a higher precision.

\begin{figure}
   %\hspace{2cm}
    \centering
 \begin{tabular}{cc}
     \includegraphics[height=2.65cm]{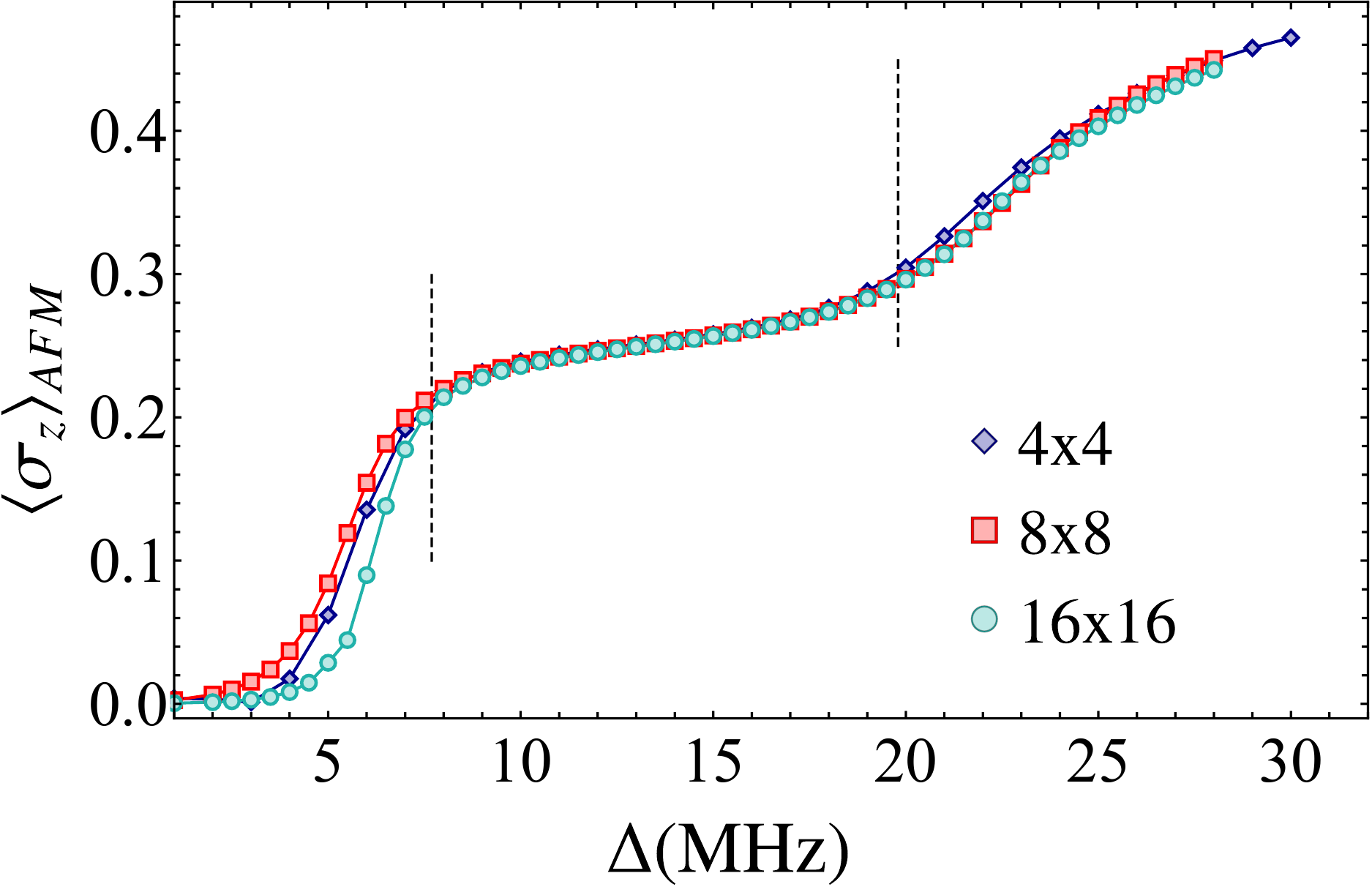} 
     & \includegraphics[height=2.65cm]{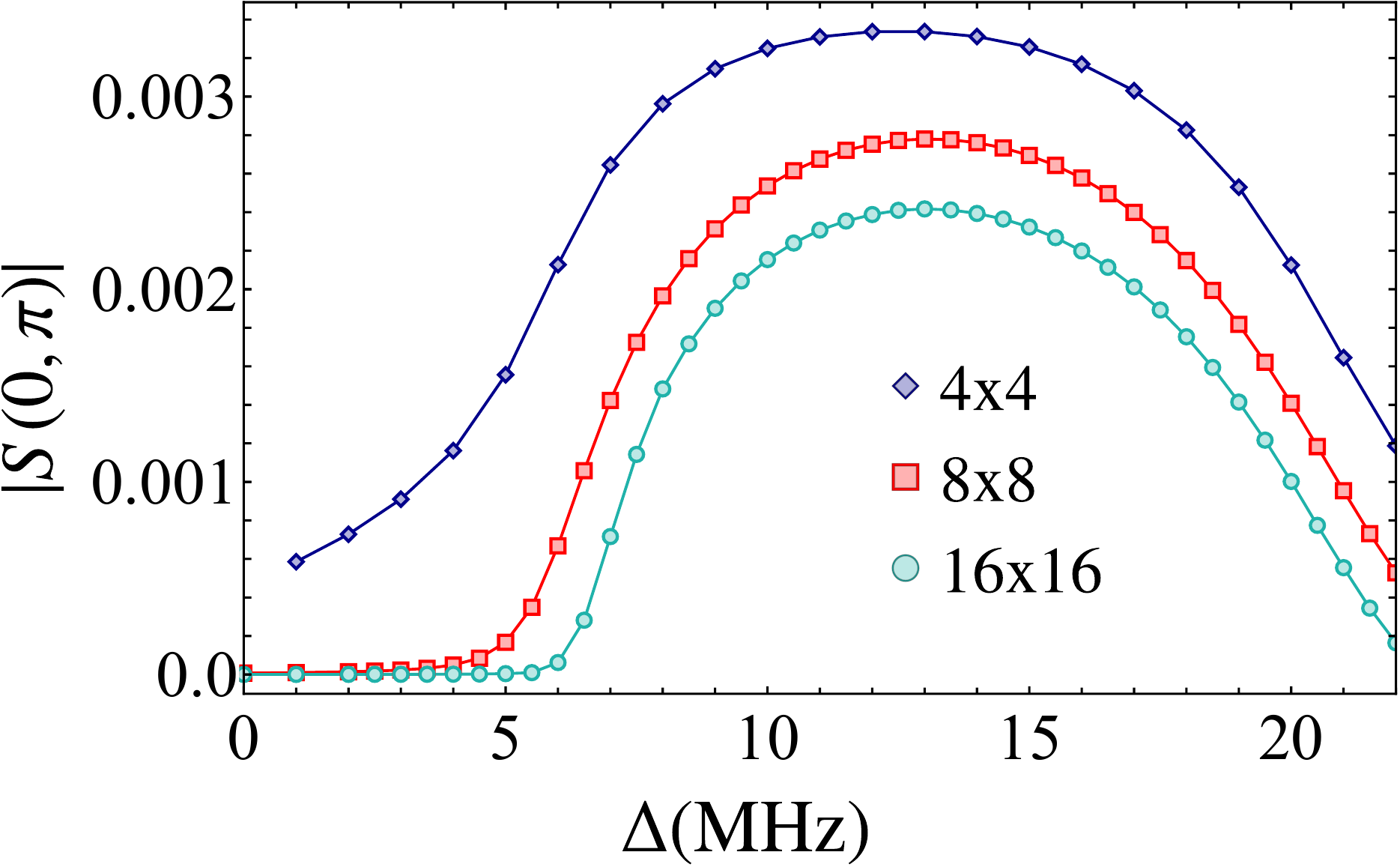} 
     \end{tabular}      \\
     \pgfputat{\pgfxy(0.4,0.0)}{\pgfbox[left,bottom]{\footnotesize (a)}}    
     \pgfputat{\pgfxy(5,0.35)}{\pgfbox[left,bottom]{\footnotesize (b)}}
      \pgfputat{\pgfxy(1.9,2.3)}{\pgfbox[left,bottom]{\footnotesize $\mathbb{Z}_4$}}       
      \pgfputat{\pgfxy(3.3,2.3)}{  \pgfbox[left,bottom]{\footnotesize $\mathbb{Z}_2$} }
 \captionsetup{justification=centerlast}
 \caption{\label{fig:clean_magn}(a) Staggered magnetization as a function of $\Delta$ at $V_{nn}=46\mathrm{MHz}$. It emerges the staircase-like behavior corresponding to the phase transitios from the disordered to the $\mathbb{Z}_2$ phase and from the  $\mathbb{Z}_2$ to the $\mathbb{Z}_4$ one. The dashed lines mark the positions of the critical points. (b) Static structure factor computed at $L=4,8,16$.
 Other parameters: $\Omega=4 \,\mathrm{MHz}$.
 }
\end{figure} 

\subsection{Rydberg atom phase diagram analysis}

In this section, we provide some detail concerning the computation of the Rydberg atom lattice phase diagram and the analysis of the phases we have observed. We focus, in particular, on the region marked by the dashed black line in Fig.~\ref{fig:phase_diagr} of the main text, corresponding to $V_{nn}=46\,\mathrm{MHz}$.

The simulations have been realized by using the aTTN ansatz with maximum bond dimension $m=300$. In order to simulate the ground state as it would be observed in an experiment, we have adopted open boundary conditions. At the same time, in order to break the degeneracy of the ground states of the $\mathbb{Z}_2$ and $\mathbb{Z}_4$ phases, we have replaced the detuning terms along the bottom and right boundary sites with the term $\Delta_{br} = -7 \,\mathrm{MHz}$, forcing the corresponding atoms not to remain in the ground state (see the panels in the phase diagram of  Fig.~\ref{fig:phase_diagr}).  For this reason, the expectation values of the correlators and the occpuations, used to estimate the structure factor and the staggered magnetization respectively, are computed only in the bulk. Moreover, we have checked that the results we obtain are in agreement with those we would obtain by imposing periodic boundary conditions.

In order to investigate the $\mathbb{Z}_2$ and $\mathbb{Z}_4$ phases, we have first computed the staggered magnetization $\langle \sigma_z\rangle_{AFM }$. Its expectation value is $1/4$ in the phase $\mathbb{Z}_4$ and $1/2$ in the $\mathbb{Z}_2$ one, in agreement with Fig~\ref{fig:clean_magn} (a), where a staircase behavior emerges as the detuning $\Delta$ is increased. The vertical dashed lines mark the positions of the disorder-$\mathbb{Z}_4$ and $\mathbb{Z}_4$-$\mathbb{Z}_2$ transitions, respectively.

\begin{figure}
   %\hspace{2cm}
    \centering
 \begin{tabular}{cc}
     \includegraphics[width=4.2cm]{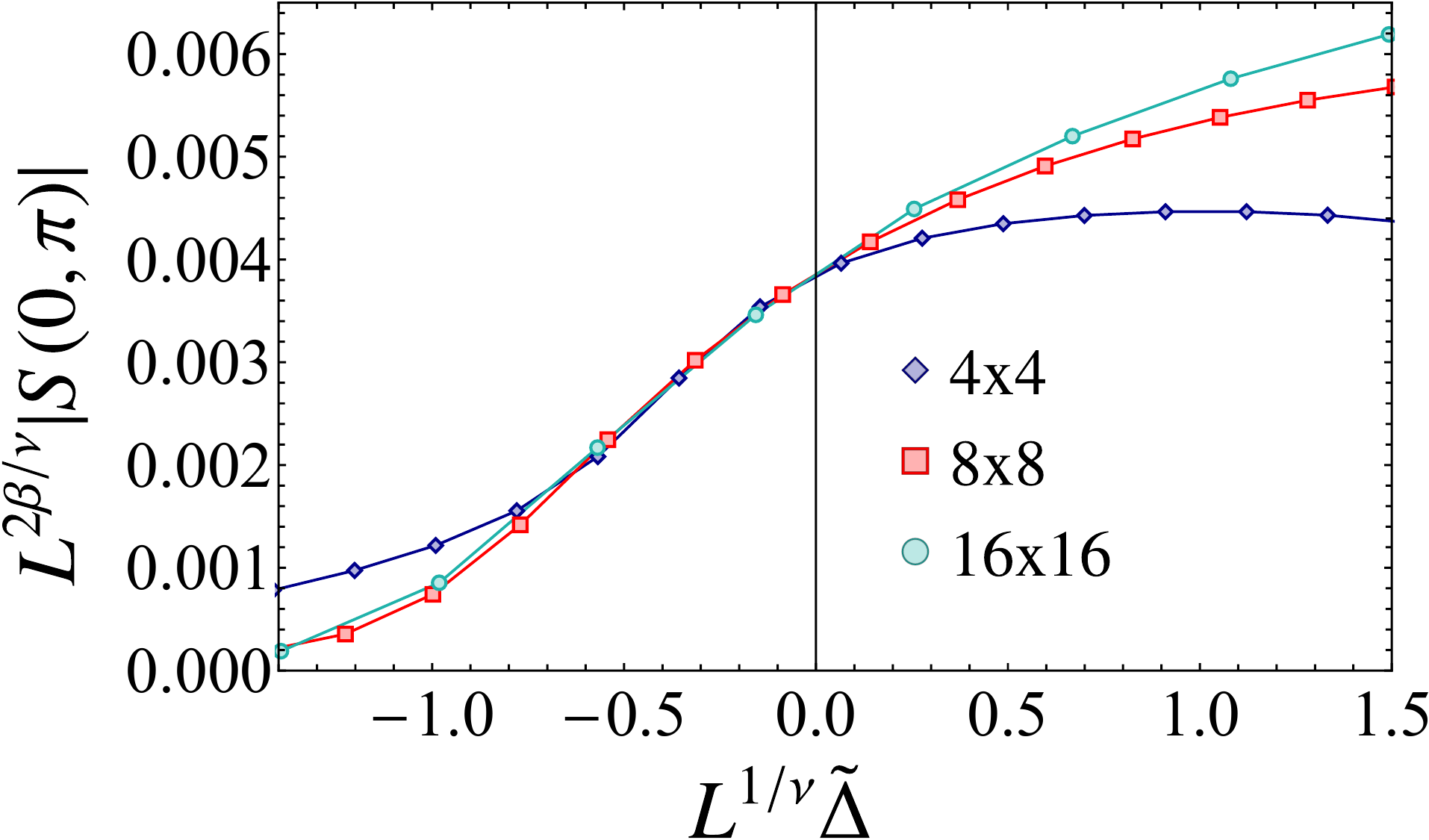} 
     & \includegraphics[width=4.2cm]{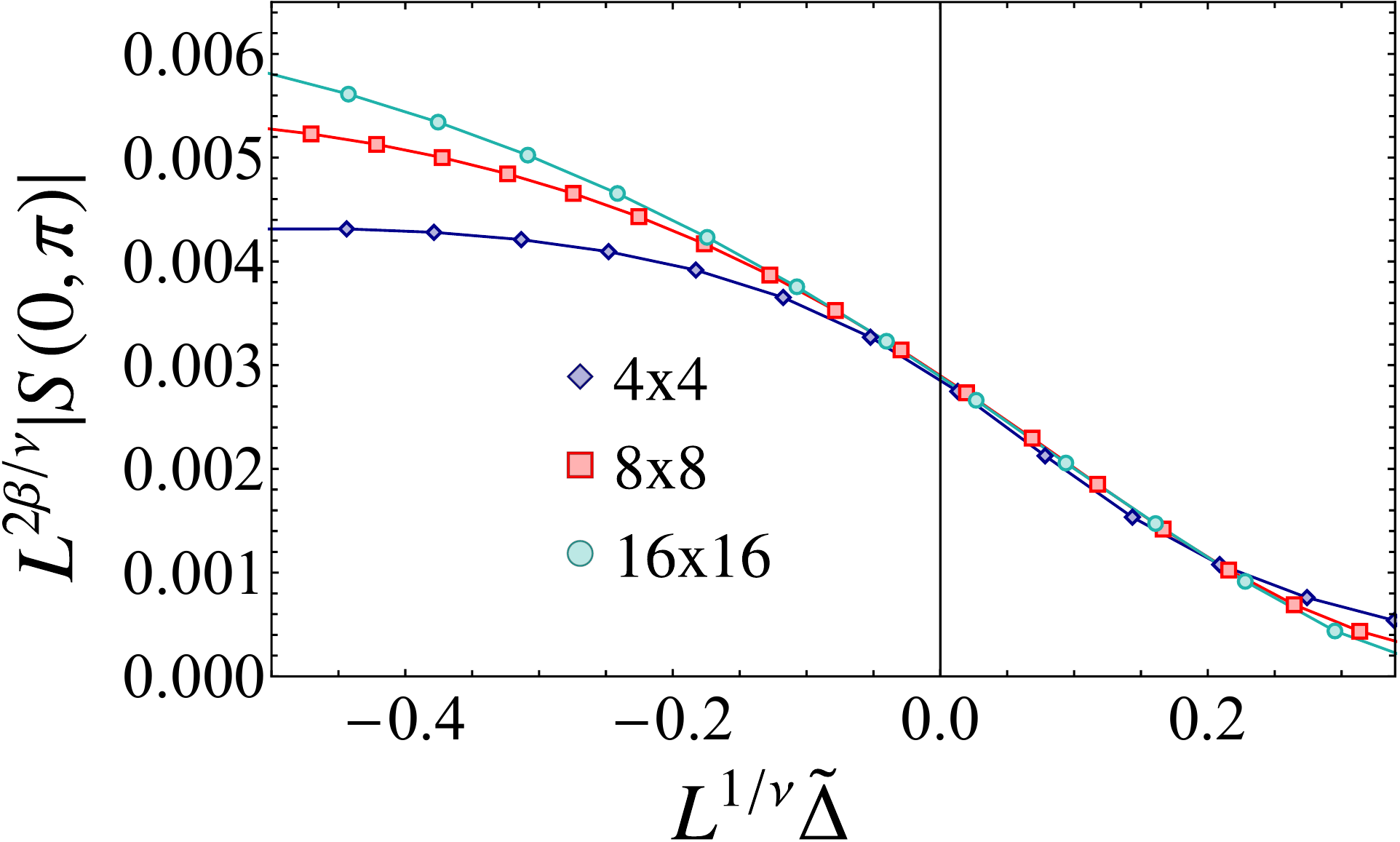} 
     \end{tabular}      \\
     \pgfputat{\pgfxy(0.8,0.1)}{\pgfbox[left,bottom]{\footnotesize (a)}}       \pgfputat{\pgfxy(5,0.45)}{\pgfbox[left,bottom]{\footnotesize (b)}}
 \captionsetup{justification=centerlast}
 \caption{\label{fig:fin_scal_an}Finite-size scaling analysis for the disorder- $\mathbb{Z}_4$ (a) and for the $\mathbb{Z}_4$-$\mathbb{Z}_2$ (b) transitions. The critical points are located at $\Delta_c=7.69\,,19.8\,,\mathrm{MHz}$ respectively.
 Other parameters: $\Omega=4 \,\mathrm{MHz}$, $V_{nn}=46\mathrm{MHz}$.
 }
\end{figure}

The positions of the critical points has been determined by performing a finite-size scaling analysis of  $\langle O^{(4)\dagger}O^{(4)}\rangle$ against the detuning $\Delta$, as they satisfy the relation \cite{Binder1981}
\begin{equation}
\langle O^{(4)\dagger}O^{(4)}\rangle L^{2\beta/\nu} = f(L^{1/\nu} \tilde{\Delta}),
\end{equation}
where $\tilde{\Delta}=(\Delta-\Delta_c)/\Delta_c$, $f(x)$ is a scaling function and $\beta,\nu$ are the critical exponents.
Since the values of $\langle O^{(4)\dagger}O^{(4)}\rangle$ coincide with those of the structure factor $S(0,\pi)$  in the disordered, $\mathbb{Z}_2$ and $\mathbb{Z}_4$ phases, we use $S(0,\pi)$ to estimate $\langle O^{(4)\dagger}O^{(4)}\rangle$ and perform the scaling analysis. As a result, we are able to determine the critical values of the detuning and the relative exponents for the two different transitions, as shown in Fig~\ref{fig:fin_scal_an}. We obtain $1/\nu=0.62(7)$ and $\beta=0.36(5)$ for the transition disorder-$\mathbb{Z}_4$ and $1/\nu=0.51(19)$ and $\beta=0.49(19)$ for the transition $\mathbb{Z}_4$-$\mathbb{Z}_2$. The errors are computed as the standard deviation of the best estimations obtained from each point on the corresponding critical lines.

Finally, we estimate the order of the transition by computing the discrete derivatives of the energy with respect to the detuning. We observe that the energy and the first derivative change continuously with the detuning, while the second derivative exhibits two peaks, confirming the presence of  two transitions (see Fig~\ref{fig:energy_an} (a)) and allowing us to conclude, therefore, that the transitions observed are of the second order. Moreover, we can use the position of the peaks to estimate the critical detuning value. We start by estimating the position of the finite-size critical point $\Delta^*(L)$ by taking the position of the peak in the corresponding curve and 
assuming that at  $\Delta^*(L)$ the correlation length is $\xi\sim L$. Since, in general, $|\Delta-\Delta_c|\sim \xi^{-\nu}$, we can estimate the critical value $\Delta^*_c$ by imposing that the linear fit for the relation $|\Delta^*(L)-\Delta^*_c|\sim L^{-\nu}$ passes through the origin. In Fig.~\ref{fig:fin_scal_an} (b) we show the results of this analysis performed for the disorder-$\mathbb{Z}_4$ transition, where we find $\Delta_c^*=8.6\pm1.3\mathrm{MHz}$ and the exponent $1/\nu=0.62$ is the one found from the finite-size scaling analysis. 

\begin{figure}
   %\hspace{2cm}
    \centering
 \begin{tabular}{cc}
     \includegraphics[height=2.58cm]{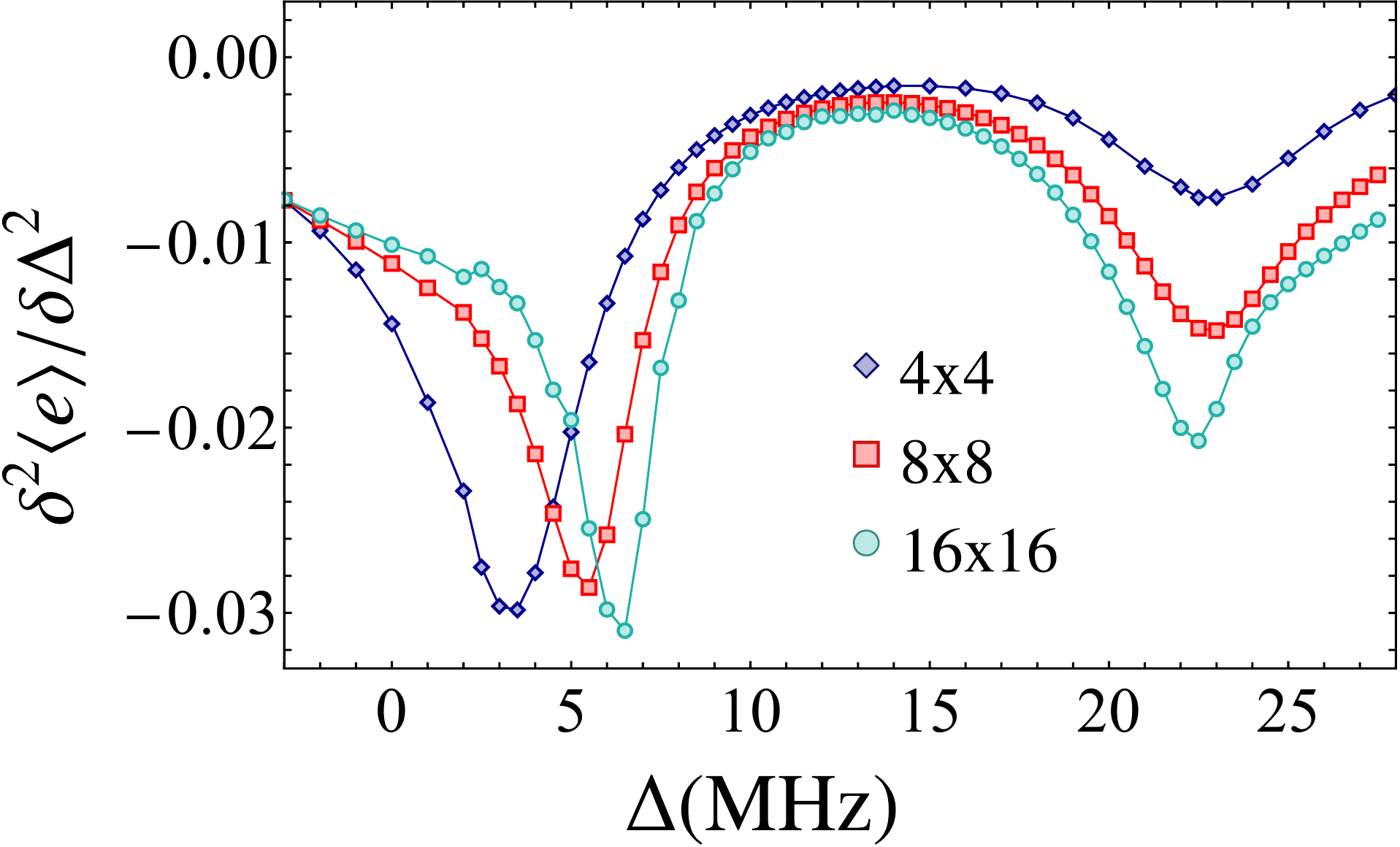} 
     & \includegraphics[height=2.59cm]{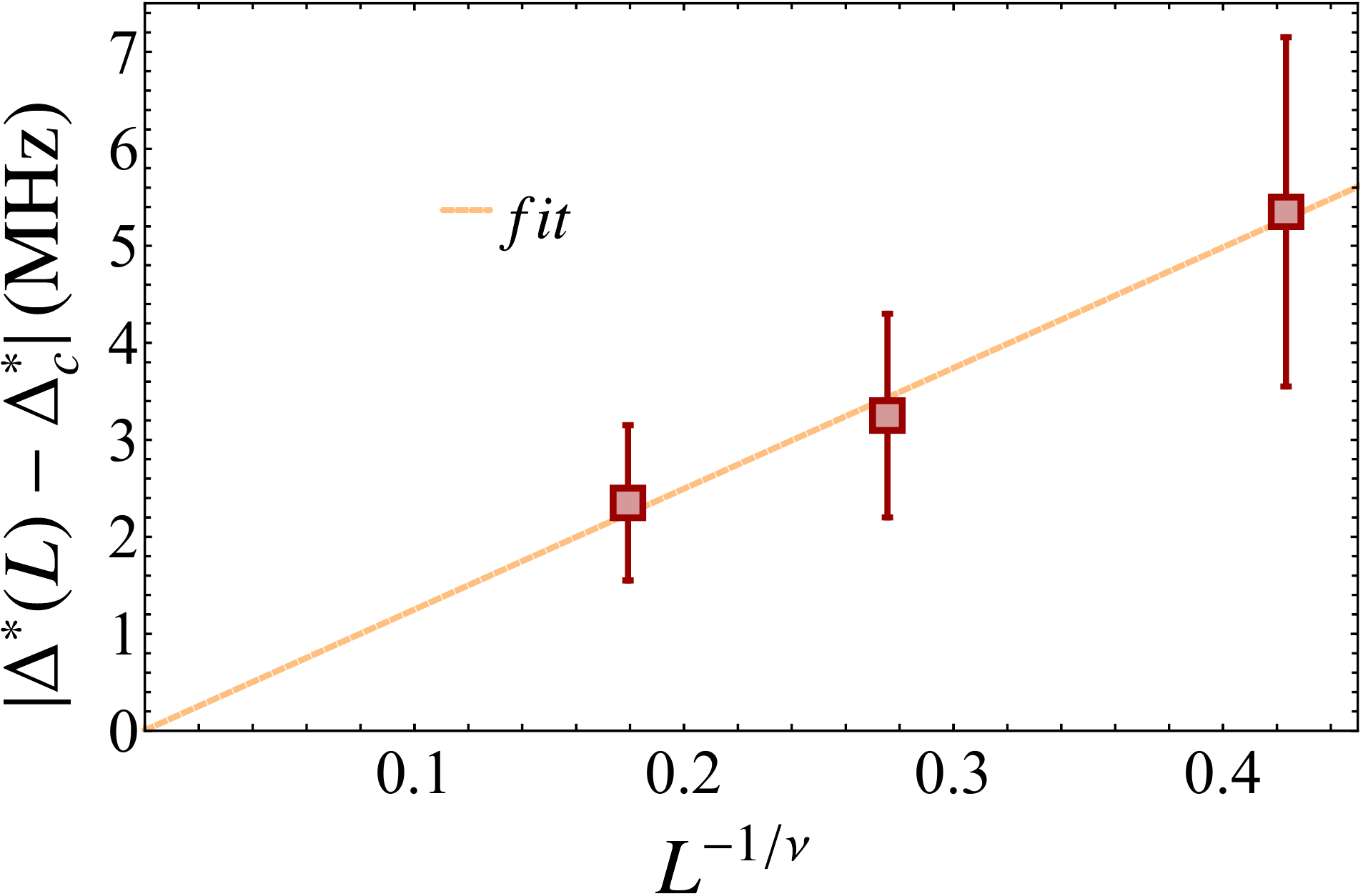} 
     \end{tabular}      \\
     \pgfputat{\pgfxy(0.8,0.1)}{\pgfbox[left,bottom]{\footnotesize (a)}}       \pgfputat{\pgfxy(5,0.45)}{\pgfbox[left,bottom]{\footnotesize (b)}}
 \captionsetup{justification=centerlast}
 \caption{\label{fig:energy_an}(a) Discrete second-order derivative of the energy, whose peaks reveals the second-order nature of the two phase transitions we observe. By using the positions of the peaks at different values of $L$ $\Delta^*(L)$ and the value of $\nu$ from the finite-size scaling analysis it is possible to obtain another estimation for the critical $\Delta_c^*=8.6\pm1.3\mathrm{MHz}$ for the disorder- $\mathbb{Z}_4$ transition.
 Other parameters: $\Omega=4 \,\mathrm{MHz}$.
 }
\end{figure} 
The error is determined by taking into account the uncertainty of the peak positions $\Delta^*(L)$. The computation of $\Delta^*_c$ allowed us to estimate the errors in the phase diagram in Fig. \ref{fig:phase_diagr} in the main text as the semi-difference between the critical values found from the finite-size scaling analysis and the energy one.

\end{document}